\newif\ifreport\reporttrue
\newif\ifblinded\blindedfalse

\ifreport
\documentclass[12pt,onecolumn,letterpaper]{article}
\else
\documentclass[10pt, conference, compsocconf]{IEEEtran}
\fi
\usepackage[usenames]{color}
\usepackage{amsmath, amssymb, xspace, epsfig, graphicx, url, cite, tabularx}
\usepackage{setspace,comment}
\usepackage{mathpartir}
\usepackage{wrapfig}
\usepackage{subfigure}
\usepackage{psfrag}
\usepackage{utf8,ttquot,times}
\usepackage{paralist}
\usepackage{multirow}
\usepackage{listings}
\usepackage[table,xcdraw]{xcolor}
\usepackage{declarations}
\usepackage{ stmaryrd }
\usepackage[font=normalsize]{caption}
\usepackage{balance}
\usepackage{tightheaders}

\ifreport
\usepackage[margin=1in]{geometry}
\date{}
\fi

\ifreport
\title{Towards a Flow- and Path-Sensitive \\ Information Flow Analysis: Technical Report}
\else
\title{Towards a Flow- and Path-Sensitive Information Flow Analysis}
\fi

\ifreport
  \author{Peixuan Li \qquad Danfeng Zhang \\ Department of Computer Science and Engineering \\
Penn State University\\ e-mail: \{pzl129,zhang\}@cse.psu.edu}
\else
\ifblinded
\author{Anonymized for Review}{}{}
\else
\author{\IEEEauthorblockN{Peixuan Li, Danfeng Zhang}
\IEEEauthorblockA{Department of Computer Science and Engineering \\ Pennsylvania State University\\ University Park, PA  United States \\
e-mail: \{pzl129,zhang\}@cse.psu.edu}}
\fi
\fi

\begin{document}
\maketitle

\begin{abstract}
This paper investigates a flow- and path-sensitive static information flow
analysis. Compared with security type systems with fixed labels, it has been
shown that flow-sensitive type systems accept more secure programs. We
show that an information flow analysis with fixed labels can be both flow- and
path-sensitive. The novel analysis has two major components: 1) a
general-purpose program transformation that removes false dataflow dependencies in a
program that confuse a fixed-label type system, and 2) a fixed-label type
system that allows security types to depend on path conditions. We formally
prove that the proposed analysis enforces a rigorous security property:
noninterference. Moreover, we show that the analysis is strictly more
precise than a classic flow-sensitive type system, and it allows sound
control of information flow in the presence of mutable variables without
resorting to run-time mechanisms.
\end{abstract}

\ifreport
\else
\fi

\section{Introduction}
\label{sec:introduction}
 
Information-flow security is a promising approach to security enforcement,
where the goal is to prevent disclosure of sensitive data by applications.
Since Denning and Denning's seminal paper~\cite{denning-cert}, static program
analysis has been widely adopted for information-flow control~\cite{sm-jsac}.
Among these program analyses, type systems (e.g.,~\cite{vsi96, myers-popl99,
ps02}) have enjoyed a great popularity due to their strong end-to-end security
guarantee, and their inherently compositional nature to combine secure
components forming a larger secure system as long as the type signatures agree.

Conventionally, we assume secrets are stored in variables, and
\emph{security levels} (e.g., $\Low$ for public and $\High$ for secret) are
associated with variables to describe the
intended secrecy of the contents. The security problem is to verify that
the final value of the public variables (outputs visible to the public) is not
influenced by the initial value of the secret variables.

Many security type systems (e.g.,~\cite{vsi96, myers-popl99, ps02}) assume
fixed levels. That is, the security level for each variable remain unchanged 
throughout program execution. Though this fixed-level assumption simplifies the
design of those type systems, one consequence is that they tend to be over-conservative (i.e., 
reject secure programs).  For example, given that
$\high$ has a level $\High$ (i.e., $\high$ holds a secret value) and $\low$ has a
level $\Low$, a fixed-level type system rejects secure programs, such as
($\low:=\high;\low:=0;$), even though the publicly observable final value of 
$\low$ is always zero.

Previous work (e.g.,~\cite{Hunt:flowsensitive}) observes that such inaccuracy
roots from the \emph{flow-insensitive} nature (i.e., the order of program
execution is ignored) of fixed-level systems. From this 
perspective, the previous example is mistakenly considered insecure because the
(impossible) execution order $(\low:=0;\low:=\high;)$ is insecure.

Hunt and Sands~\cite{Hunt:flowsensitive} propose a classic flow-sensitive
type system which allows a variable to have multiple security levels
over the course of computation. For example, this floating-level type system correctly
accepts the program ($\low:=\high;\low:=0;$) by assigning $\low$ with levels
$\High$ and $\Low$ after the first and second assignments respectively.
However, this floating-level system is still path-insensitive, meaning that the
predicates at conditional branches are ignored in the analysis. For example,
it incorrectly rejects the following secure program since the (impossible) branch combination 
$(y:=\high;\low:=y;)$ is insecure.
\begin{lstlisting}[numbers=none,xleftmargin=5ex]
$\ifcmd{x=1}{y:=0}{y:=\high};$
$\ifthen{x=1}{\low:=y}$
\end{lstlisting}

This paper develops a flow- and path-sensitive information flow
analysis that is precise enough to accept the aforementioned secure programs.
The novel analysis is built on two key observations. First,
flow-sensitivity can be gained via a general-purpose program transformation
that eliminates false dataflow dependencies that confuse a flow-insensitive 
type system. Consider the example ($\low:=\high;\low:=0;$) again. The 
transformation removes
the false dataflow dependency between $\high$ and $\low$ by introducing an
extra copy of the variable $\low$ and keeps track of the \emph{final
copy} of each variable at the same time. So, the example is transformed to
($\low_1:=\high;\low_2:=0;$), where $\low_2$ is marked as the final copy. Then,
a fixed-level system can easily type-check this program by assigning
levels $ \High $ and $ \Low $ to $ \low_1 $ and $ \low_2 $ respectively. 

Second, path-sensitivity can be gained via consolidating dependent type theory
(e.g., \cite{Xi:POPL99, Swamy:fable, Condit:deputy}) into security labels. That is,
a \emph{security label} is, in general, a function from program states to security
levels.
Consider the second example above with branches. We can assign $y$ a 
dependent security label: $(x=1?\Low:\High)$, meaning that the level of $y$ is
$\Low$ when $x=1$, and $\High$ otherwise.  Hence, the information flow from $y$
to $\low$ can be judged as secure since it only occurs when $x=1$ (hence, $y$
has level $\low$).

Based on the key observations, we propose a flow- and path-sensitive
information flow analysis that consists of two major components: a general
purpose program transformation that removes false dataflow dependencies
that otherwise compromise the precision of a fixed-level system, as well as a fixed-label type
system with dependent labels. Each component of our analysis targets one 
insensitive source of
previous type systems. The modular design not only enables tunable precision of
our analysis, but also sheds light on the design of security type systems: we 
show that a fixed-level system (e.g.,~\cite{vsi96}) plus the program
transformation is \emph{as precise as}\footnote{We note that in the
information flow literature, different terms (such as ``precision'' and
``permissiveness'') have been used to compare the amount of false positives of
various mechanisms~\cite{Nataliia16}. In this paper, we say a static analysis
A is \emph{as precise as} a static analysis B if A accepts every
secure program that is accepted by B. Moreover, we say A is \emph{(strictly)
more precise than} B if A is as precise as B, and A accepts at least one secure
program that is rejected by B.} the classic flow-sensitive system
in~\cite{Hunt:flowsensitive}; furthermore, a fixed-label dependent type system 
can soundly control information flow in the presence of mutable variables
without resorting to run-time mechanisms (e.g.,~\cite{hwtiming,trustzone}).

This paper makes the following key contributions:

\begin{enumerate}[1)]
\item We formalize a novel flow- and path-sensitive information flow analysis
for a simple WHILE language. The analysis consists of a novel program
transformation, which eliminates imprecision due to flow-insensitivity
(Section~\ref{sec:transformation}), and a purely static type system using
dependent security labels (Section~\ref{sec:typesystem}).

\item We formally prove the soundness of our analysis
(Section~\ref{sec:soundness}): the source program satisfies
termination-insensitive noninterference whenever the transformed program
type-checks. Novel proof techniques are required due to the extra variables
introduced (for added precision) in the transformed program.

\item We show that our analysis is strictly more precise than a classic
flow-sensitive type system~\cite{Hunt:flowsensitive}
(Section~\ref{sec:comparison}). One interesting consequence is that the program
transformation automatically makes a sound flow-insensitive type system
(e.g.,~\cite{vsi96}) as precise as the classic flow-sensitive
system~\cite{Hunt:flowsensitive}.

\item We show that our dependent type system soundly controls information flow
in the presence of mutable variables without resorting to dynamic mechanisms,
such as the dynamic erasure mechanism in previous work~\cite{hwtiming,trustzone}.

\end{enumerate}

\section{Background and Overview}

\newsavebox\FlowExample
\begin{lrbox}{\FlowExample}
\ifreport
	\begin{minipage}{0.25\textwidth}
\else
	\begin{minipage}{0.18\textwidth}
\fi
		\begin{lstlisting}[numbers=left,xleftmargin=15pt,framexleftmargin=15pt]
$x := \high$; 
$\bkassign{x}{0};$ 
$\low := x$;
		\end{lstlisting}
	\end{minipage}
\end{lrbox}

\newsavebox\FlowExampleTrans
\begin{lrbox}{\FlowExampleTrans}
\ifreport
	\begin{minipage}{0.25\textwidth}
\else
	\begin{minipage}{0.18\textwidth}
\fi
		\begin{lstlisting}[numbers=left,xleftmargin=15pt,framexleftmargin=15pt]
$x := \high$; 
$\assign{x_1}{0};$ 
$\low := x_1$;
		\end{lstlisting}
	\end{minipage}
\end{lrbox}

\newsavebox\PathExample
\begin{lrbox}{\PathExample}
	\begin{minipage}{0.35\textwidth}
		\begin{lstlisting}[numbers=left,xleftmargin=15pt,framexleftmargin=15pt]
$x:=0$;  $y:=0$; 
$\If$ ($\low_1$ < 0) $\Then$ $y := \high$; 
$\If$ ($\low_1$ > 0) $\Then$ $x := y$; 
$\low_2 := x$;
		\end{lstlisting}
	\end{minipage}
\end{lrbox}

\begin{figure*}
	\centering
	\subfigure[Flow-Insensitive Analysis Rejects Secure Program.]{
		\label{fig:example:flow}\
		\usebox\FlowExample
	}
	\quad
	\subfigure[Flow-Insensitive Analysis Accepts Equivalent Program.]{
		\label{fig:example:flow:trans}\
		\usebox\FlowExampleTrans
	}
	\quad
	\subfigure[Path-Insensitive Analysis Rejects Secure Program.]{
		\label{fig:example:path}\
		\usebox\PathExample
	}
	\caption{Examples: Imprecise Information Flow Analysis Rejects Secure Programs.}
	\label{fig:example}
\end{figure*}

\subsection{Information Flow Analysis}

We first review standard information flow terminology used in this
paper. We assume all variables are associated with security levels. A security
policy is specified as the ordering of the security levels, typically in the
form of a security lattice. For data $d_1$ with security level $\ell_1$ and
data $d_2$ with level $\ell_2$, the policy allows information flow from $d_1$
to $d_2$ if and only if $\ell_1\absleq \ell_2$. In this paper, we use two
distinguished security levels \High (Secret) and \Low (Public) for simplicity,
but keep in mind that the proposed theory is general enough to express richer security levels.
The security policy on the levels $\Low$ and $\High$ is defined as $\Low\absleq\High$,
while $\High\not\absleq\Low$. That is, information flow from public data to 
secret variable
is allowed, while the other direction is forbidden.  Hereafter, we assume
variable \high is labeled as \High, and variable \low is labeled as \Low unless
specified otherwise.

\paragraph{Explicit and Implicit Flows}
An information flow analysis prohibits any explicit or implicit information
flow that is inconsistent with the given policy. \emph{Explicit flows} take 
place when  confidential data are passed directly to public variables, such as 
the
command $\low := \high$, while \emph{implicit flows} arise from the control
structure of the program. For example, the following program has an implicit
flow: 
\[\ifcmd{\high=0}{\low:=0}{\low:=1}\]

Assume the secret variable \high is either 0 or 1. This code is insecure since it is
functionally equivalent to $\low := \high$. That is, the confidential data \high
is copied to a public variable $\low$.

An information flow security system rules out all explicit and
implicit flows; any violation of a given security policy 
results in an
error. As in most information flow analyses, we do not consider timing,
termination and other side channels in this paper; controlling side channel
leakage (e.g.,~\cite{agat00,Kopf:Durmuth:CSF2009,zam12}) is largely an
orthogonal issue.

\subsection{Sources of Imprecision}
\label{sec:imprecision}

Most information flow analyses provide soundness (i.e., if the analysis
determines that a program is secure, then the program provably prevents
disclosure of sensitive data). However, since the problem of checking
information flow security is in general undecidable~\cite{sm-jsac}, one key
challenge of designing an information flow analysis is to maintain soundness,
while improving precision (i.e., reject fewer secure programs). 

In this section, we introduce the major sources of imprecision in existing type
systems. In the next section (Section~\ref{sec:overview}), we illustrate how
does our novel information flow analysis alleviate those sources of
imprecision.

\paragraph{Flow-Insensitivity} The first source of imprecision is
\emph{flow-insensitivity}, meaning that the order of execution is not taken
into account in a program analysis~\cite{nielson2015principles}. In the context 
of information flow analysis, the intuition is that an analysis is 
flow-insensitive if a program is analyzed as secure only when every
subprogram is analyzed as secure~\cite{Hunt:flowsensitive}. 

Many security type systems, including~\cite{vsi96, myers-popl99, ps02}, are 
flow-insensitive. Consider
the program in Figure~\ref{fig:example:flow} (for now, ignore the brackets). This program is secure since the 
public variable $\low$ has a final value zero
regardless of the secret variable \high.  However, it is considered insecure by a
flow-insensitive analysis because of the insecure subprogram $(x := \high; \low 
:= x;)$. 
Under the hood, the imprecision arises since the analysis
requires fixed levels: the security level of a variable must remain the same
throughout the program execution. But in this example, these is no fixed-level
for the variable $x$: when the level is \High, $\low:=x$ is insecure;
when the level is \Low, $x:=\high$ is insecure.

\paragraph{Path-Insensitivity}
The second source of imprecision is \emph{path-insensitivity}, meaning that the
predicates at conditional branches are ignored in a program
analysis~\cite{nielson2015principles}. In the context of information flow
analysis, the intuition is that an analysis is path-insensitive if a program is
analyzed as secure only when every sequential program generated from one
combination of branch outcomes is analyzed as secure. 

For instance, the flow-sensitive type system in~\cite{Hunt:flowsensitive} is
path-insensitive; consequently, it rejects the secure program shown in
Figure~\ref{fig:example:path} (due to Le Guernic and
Jensen~\cite{Guernic:monitor}). This example is secure since the value of the
secret variable \high never flows to the public variable $\low_2$, since the
assignments $y:=\high$ and $x:=y$ never execute together in the same program execution.
However, the type system in~\cite{Hunt:flowsensitive} rejects this program
because it lacks the knowledge that the two if-statements cannot take the
``then'' branch in the same execution.  Hence, it has to conservatively analyze
the security of an impossible program execution:
$x:=0;y:=0;y:=\high;x:=y;\low:=x$, which is insecure due to an explicit flow
from $\high$ to $\low$. 

Under the hood, we observe that the imprecision arises from the fact that a
path-insensitive analysis (e.g.,~\cite{Hunt:flowsensitive}) requires that the
security levels of a variable on two paths to be ``merged'' (as the least
upper bound) after a branch. Consider the first branch in
Figure~\ref{fig:example:path}. The ``then'' branch requires $y$ to be \High due
to the flow from \high to $y$.  So after that if-statement, the label of $y$
must be \High (i.e., which path is taken is unknown to the rest of the
program).  Similarly, $x$ has label \High after the second if-statement. Hence,
$\low_2:=x$ is rejected due to an explicit flow from \High to \Low.

\subsection{Overview}
\label{sec:overview}

In order to alleviate analysis imprecision due to 
flow- and path-insensitivity, our novel information flow analysis has two major
components: a program transformation that enables flow-sensitivity and a type
system with dependent security labels, which enables path-sensitivity.

\subsubsection{Program Transformation} 
Consider the example in Figure~\ref{fig:example:flow} (for now, ignore the
brackets). A fixed-level type
system rejects this program since the levels of $x$ at line 1 and 3
are inconsistent. We observe that there are indeed two \emph{copies} of
 $x$ in this program but only the final one (defined at line 2) is
released. So without modifying a type system, we can explicitly transform the
source program to a semantically equivalent one that explicitly marks different
copies.

The source language of our program analysis (Section~\ref{sec:language})
provides a tunable knob for improved precision: a \emph{bracketed assignment}
in the form of $\bkcmd{\assign{x}{e}}$. Such an assignment is semantically
identical to $\assign{x}{e}$ but allows a programmer to request improved
precision (the source language allows such flexibility since reduced precision
might be preferred for reasons such as more efficient analysis on the program).
In particular, for a bracketed assignment $\bkcmd{\assign{x}{e}}$, the
program transformation (Section~\ref{sec:transformation}) generates a fresh
copy for $x$ and uses that copy in the rest of program until another new copy is
generated.  For example, given the bracketed assignment at line 2 of
Figure~\ref{fig:example:flow}, the transformed program is shown in
Figure~\ref{fig:example:flow:trans}, where the second definition of $x$ and its
use at line 3 are replaced with $x_1$. The benefit is that 
the \emph{false dataflow dependency} from $\high$ to $\low$ in the
source program is eliminated. Hence, the transformed program can be accepted by
a fixed-level type system, by assigning $x$ and $x_1$ to levels $\High$ and
$\Low$ respectively. In general, we prove that (when all assignments are
bracketed) the transformation enables a fixed-level system to be at least as
precise as a classic flow-sensitive type system
(Section~\ref{sec:comparison}).


\subsubsection{Dependent Labels} 
Consider the example in Figure~\ref{fig:example:path}. A path-insensitive type
system rejects this program since such a type system ignores the path
conditions under which assignments occur. Consequently, the security level of
$y$ is conservatively estimated as $\High$ after line 2, though when
$\low_1≥0$, variable $y$ only carries public information.

In our system, path-sensitivity is gained via dependent security labels (i.e.,
security labels that depend on program states). Compared with a security
level drawn directly from a lattice, a dependent security label precisely
tracks all possible security levels from different branches; hence,
path-sensitivity is gained. Since dependent security labels are orthogonal to
bracketed assignments, extra precision can be gained in our system even in the
absence of bracketed assignments.
For example, while the program in Figure~\ref{fig:example:path} can not be
accepted using any simple security level for $y$, we can assign to $y$ a dependent
label $(\low_1<0?\High:\Low)$, which specifies an invariant that the level of $y$ is
$\High$ when $\low_1<0$ (i.e., the ``then'' branch is taken at line 2); the
level is $\Low$ otherwise. Such an invariant can be maintained by the type system
described in Section~\ref{sec:typesystem}. For instance, to ensure that the
explicit flow from $y$ to $x$ at line 3 is
secure, the type system generates a proof obligation $(\low_1\! >\! 0 ⇒
(\low_1\!<\!0?\High\!:\!\Low) ⊑ \Low)$, meaning that the information flow from $y$ to
$x$ must be permissible under the path condition $\low_1<0$.  This proof
obligation can easily be discharged by an external solver.  The soundness of
our type system (Section~\ref{sec:soundness}) guarantees that all security
violations are detected at compile time.

\section{Language Syntax and Semantics}
\label{sec:language}

\begin{figure}
\begin{align*}
	&\text{Vars}       &x, y, z ∈\; &\Vars \\
	&\text{Expr} \quad &\expr ::=\; &x \mid n \mid \expr~\op~\expr \\
	&\text{Cmds} \quad &\cmd  ::=\; &\Skip \mid \cmd_1;\cmd_2 \mid \assign{x}{\expr} \mid \bkassign{x}{e} \mid\\
	&                  &            &\ifcmd{\expr}{\cmd_1}{\cmd_2} \mid \while{\expr}{\cmd}
\end{align*}
\caption{Syntax of the Source Language.}
\label{fig:while-syntax}
\end{figure}

\begin{figure*}
\begin{mathpar}
\inferrule{ }{\configTwo{\Mem}{n} \evalto {n}} 
\and
\inferrule{ }{\configTwo{\Mem}{x} \evalto {\Mem(x)}}
\and
\inferrule{\inferrule{}{\configTwo{\Mem}{e_1} \evalto {n_1} \\\\
	\configTwo{\Mem}{e_2} \evalto {n_2}} \and \
  n = n_1\ \op\ n_2}
 {
  \configTwo{\Mem}{e_1\ \op\ e_2} \evalto {n}}
\end{mathpar}

\begin{mathpar}
\inferrule[S-Skip]
{ }{\configTwo{\Mem}{\Skip;\cmd} \To \configTwo{\Mem}{\cmd}}
\and
\inferrule[S-Assign]
{\configTwo{\Mem}{e} \evalto {n} }
{\configTwo{\Mem}{\assign{x}{e}} \To \configTwo{\Mem\Mupdate{x}{n}}{\Skip} }
\and
\inferrule[S-Assign-Bracket]
{\configTwo{\Mem}{e} \evalto {n} }
{\configTwo{\Mem}{\bkassign{x}{e}} \To \configTwo{\Mem\Mupdate{x}{n}}{\Skip} }
\and
\inferrule[S-Seq]
{\configTwo{\Mem}{\cmd_1} \To \configTwo{\Mem'}{\cmd_1'} }
{\configTwo{\Mem}{\cmd_1; \cmd_2} \To  \configTwo{\Mem'}{\cmd_1';\cmd_2}}
\and
\inferrule[S-While]
{ }
{\configTwo{\Mem}{\while{e}{\cmd}} \To \configTwo{\Mem}{\ifcmd{e}{(\cmd;\while{e}{\cmd})}{\Skip}} }
\and
\inferrule[S-If1]
{\configTwo{\Mem}{e} \evalto n \quad n \neq 0 }
{\configTwo{\Mem}{\ifcmd{e}{\cmd_1}{\cmd_2}} \To \configTwo{\Mem}{\cmd_1}}
\and
\inferrule[S-If2]
{\configTwo{\Mem}{e} \evalto n \quad n = 0 }
{\configTwo{\Mem}{\ifcmd{e}{\cmd_1}{\cmd_2}} \To \configTwo{\Mem}{\cmd_2}}

\end{mathpar}

%
\caption{Semantics of the Source Language.}
\label{fig:while-semantics}
\end{figure*}

In this paper, we consider a simple imperative WHILE language whose syntax and
operational semantics are shown in Figures~\ref{fig:while-syntax}
and~\ref{fig:while-semantics} respectively. The syntax and semantics are
mostly standard: expressions $ e $ consist of variables $ x $, integers $ n $,
and composed expressions $ e ~\op~ e$, where $ \op $ is a binary arithmetic
operation. Commands $ \cmd $ consist of standard imperative instructions, including
$\Skip $, sequential composition $c_1;c_2$, assignments, conditional $ \If $
branch and $ \While $ loop.  The semantics of expressions are given in the form
of $ \configTwo{\Mem}{e} \evalto n $ (big-step semantics), where memory $m$ 
maps variables to their values. The small-step semantics of
commands has the form of $ \configTwo{\Mem}{c}→\configTwo{\Mem'}{c'}$, where $ \configTwo{\Mem}{c} $ is a
configuration. We use $ \Mem\Mupdate{x}{n} $ to denote the memory that is
identical to $\Mem$ except that variable $ x $ is updated to the new value $n$.

The only interesting case is the bracketed assignment $ \bkassign{x}{e} $,
which is semantically equivalent to normal assignment $ \assign{x}{e} $ in the
source language. These commands are tunable knobs for improved precision in
our information flow analysis, as we show shortly.

\section{Program Transformation}
\label{sec:transformation}

To alleviate the imprecision due to flow-insensitivity, one component of
our analysis is a novel program transformation that introduces extra variable copies
to the source program, so that false dataflow dependencies that otherwise may
confuse flow-insensitive analyses are removed.

\subsection{Bracketed Assignments and the Transformed Program}
We propose a general and flexible design for the program transformation. In
particular, the program transformation is triggered only for
assignments that are marked with brackets. Such a design enables a tunable
control of analysis precision for programmers or high-level program analysis
built on our meta source language: when there is no bracketed assignment, the
transformed program is simply identical to the source program; when all
assignments have brackets, the transformation generates a fresh copy of $x$ for
each bracketed assignment $\bkassign{x}{e}$.

Due to the nature of the transformation, the transformed program follows the
same syntax and semantics as the source language, except that  all bracketed
assignments are removed.

To avoid confusion, we use underlined notations for the transformed program: 
$\et$ for expressions, $ \cmdt $ for commands and $ \Mt $ for memories, when
both the original and the transformed programs are in the context; otherwise,
we simply use $\expr$, $\cmd$ and $\Mem$ for the transformed programs as well.

\subsection{Transformation Rules}
\begin{figure*}
\begin{mathpar}
	\inferrule{}{\configTwo{\actset}{n} \tto  {n}} 
	\and 
	\inferrule{}{\configTwo{\actset}{x} \tto {\actset(x)}}
	\and
	\inferrule{\configTwo{\actset}{e_1} \tto {\et_1'}
				\quad \configTwo{\actset}{e_2} \tto {\et_2'} 	}
	{\configTwo{\actset}{e_1\ \op\ e_2} \tto \et_1'\ \op\ \et_2'}
\end{mathpar}
\begin{mathpar}
\inferrule[TRSF-Skip]
{}{\configTwo{\actset}{\Skip} \tto \configTwo{\actset}{\Skip} }
\and
\inferrule[TRSF-Assign]
{\configTwo{\actset}{e} \tto \et }
{\configTwo{\actset}{\assign{x}{e}} \tto \configTwo{\actset\Mupdate{x}{x}}{\assign{x}{\et}} }
\and
\inferrule[TRSF-Assign-Create]
{\configTwo{\actset}{e} \tto \et 
	\quad  i \text{ is a fresh index for } x}
{\configTwo{\actset}{\bkassign{x}{e}} \tto \configTwo{\actset\Mupdate{x}{x_i}}{\assign{x_i}{\et}} }
\and
\inferrule[TRSF-Seq]
{\configTwo{\actset}{\cmd_1} \tto \configTwo{\actset_1}{\cmdt_1} 
	\quad \configTwo{\actset_1}{\cmd_2} \tto \configTwo{\actset_2}{\cmdt_2}}
{\configTwo{\actset}{\cmd_1; \cmd_2} \tto \configTwo{\actset_2}{\cmdt_1;\cmdt_2}}
\and
\inferrule[TRSF-If]
{\configTwo{\actset}{e} \tto \et 
	\quad \configTwo{\actset}{\cmd_1} \tto \configTwo{\actset_1}{\cmdt_1} 
	\quad \configTwo{\actset}{\cmd_2} \tto \configTwo{\actset_2}{\cmdt_2} 
	\quad \Phi(\actset_1, \actset_2)  \tto \actset_3 }
{ \configTwo{\actset}{\ifcmd{e}{\cmd_1}{\cmd_2}} \tto 
\configTwo{\actset_3}{\ifcmd{\et}{(\cmdt_1;\phiassigneta{\actset_3}{\actset_1})}{(\cmdt_2;\phiassigneta{\actset_3}{\actset_2
 })} } }
\and
\inferrule[TRSF-While]
{ \configTwo{\actset}{\cmd} \tto \configTwo{\actset_1}{\cmdt_1} 
	\quad \Phi(\actset, \actset_1)  \tto \actset_2 
	\quad \configTwo{\actset_2}{\cmd} \tto \configTwo{\actset_3}{\cmdt} 
	\quad \configTwo{\actset_2}{e} \tto \et  }
{ \configTwo{\actset}{\while{e}{\cmd}} \tto \configTwo{\actset_2}{\phiassigneta{\actset_2}{\actset};\while{\et}{(\cmdt;\phiassigneta{\actset_2}{\actset_3}) }} }

\end{mathpar} 
\caption{Program Transformation. We use $\assign{\actset}{\actset'}$ as a shorthand for $\{\assign{\actset(v)}{\actset'(v)} \mid v\in \Vars \AND \actset(v)\not=\actset'(v)\}$.}
\label{fig:transformation}     
\end{figure*}

\begin{figure}	
	\begin{mathpar}
		\mergefunc(\actset_1, \actset_2) = λx.
		\begin{cases}
			x_i, ~i \text{ fresh for } x, &\! \actset_1(x)\! \not=\!  \actset_2(x)\\
			\actset_1(x), &\! \actset_1(x)\! =\!  \actset_2(x) \\
		\end{cases}
	\end{mathpar}
	\[	
	\inferrule*[right=TRSF-Phi]
	{ \actset_3=\mergefunc(\actset_1, \actset_2) }
	{ \Phi(\actset_1, \actset_2)  \tto {\actset_3} 	}	
	\]
	\caption{Merge Function.}
	\label{fig:mergephi}
\end{figure}

The program transformation maintains one \emph{active copy} for each variable
in the source code. One invariant maintained by the transformation is that
for each program point, there is \emph{exactly one active copy} for each
source-program variable. Intuitively, that unique active copy holds the most
recent value of the corresponding source-program variable.

\begin{Definition}[Active Set]
An active set $ \actset:\Vars \mapsto \underline{\Vars} $, is an
\emph{injective function} that maps a source variable to a unique variable in
the transformed program.
\end{Definition}

For simplicity, we assume that the variables in the transformed program follow the
naming convention of $x_i$ where $x\in \Vars$ and $i$ is an index. Hence, for
any variable $\vt$ in the range of $\actset$, we simply use $\proj{\vt}$ to denote
its corresponding source variable (i.e., a variable without the index). Hence, $\vt =
\actset(\proj{\vt})$ always holds by definition.  Moreover, since we frequently
refer to the range of $\actset$, we abuse the notation of $\actset$ to denote
active copies that $\actset$ may map to (i.e., the range of $\actset$). That
is, we simply write $\vt\in \actset$ instead of $\vt\in \Range(\actset)$ in this
paper. Moreover, we use $ \actset\Mupdate{x}{x_i} $ to denote an active set
that is identical to $\actset$ except that $x$ is mapped to $x_i$.

The transformation rules are summarized in Figure~\ref{fig:transformation}.
For an expression $e$, the transformation has the form of
$\configTwo{\actset}{e} \tto \et$, where $\et$ is the transformed expression.
The transformation of an expression simply replaces the source variables with
their active copies in $\actset$. 

For a command $c$, the transformation has the form of
$\configTwo{\actset}{\cmd} \tto \configTwo{\actset'}{\cmdt}$, where $c$ is the
source command and $\cmdt$ is the transformed one. Since assignments may update
the active set, $\actset'$ represents the active set after $\cmdt$.  

Rule (TRSF-Assign) applies to a normal assignment. It transforms the assignment to one
with the same assignee and update $\actset$ accordingly. Rule (TRSF-Assgin-Create) applies to a bracketed
assignment $ \bkassign{x}{e} $. It renames the assignee to a fresh variable. For
example, line 1 of  the transformed program in
Figure~\ref{fig:example:flow:trans} is exactly the same as the original program
in \ref{fig:example:flow}; but the assignee of line 2 is renamed to $x_1$.
Rule~\ruleref{Trsf-If} uses a special $ \Phi $ function, defined in
Figure~\ref{fig:mergephi}, to merge the active sets generated from the
branches. In particular, $\Phi(\actset_1, \actset_2)\tto \actset_3 $ generates an active set
$ \actset_3 $ that maps $x$ to a fresh variable iff $\actset_1(x)\not=\actset_2(x)$. 
Transformation for the $ \While $ loop is a little tricky since we need to compute
an active set that is active both before and after each iteration.
Rule~\ruleref{TRSF-While} shows one feasible approach: the rule transforms the
loop in a way that $\actset_1$ is a fixed-point: the active set is always
$\actset_1$ before and after an iteration by the transformation. 

We note that given an identity function as the initial active set
$\actset$, a program without any bracketed assignment is transformed to itself
with a final active set $\actset$. At the other extreme, the transformation
generates one fresh active copy for each assignment when all assignments are
bracketed.

\subsection{Correctness of the Transformation}
One important property of the proposed transformation is its correctness: a
transformed program is semantically equivalent to the source program. To
formalize this property, we need to build an equivalence relation on the memory
for the source program ($\Mem: \Vars→\nat$) and the memory for the transformed
program ($\Mt: \underline{\Vars}→\nat$). We note that the projection of $\Mt$
on an active set $\actset$ defined as follows shares the same domain and range
as $\Mem$. Hence, it naturally specifies an equivalence relation on $\Mem$ and
$\Mt$ w.r.t.  $\actset$: $\Mem$ can be directly compared with $\Mt^{\actset}$.

\begin{Definition}[Memory Projection on Active Set]
\label{def:memactset}
We use $ \Mt^{\actset} $ to denote the projection of $\Mt$ on the active set $\actset$, defined as follows:
	\begin{align*}
	~ 
		 & ∀x\in \Vars.~\Mt^{\actset}(x) = \Mt(\actset(x))
	\end{align*}
\end{Definition}

We formalize the correctness of our transformation as the following theorem.
As stated in the theorem, the correctness is not restricted to any particular
initial active set $\actset$. 
 
\begin{Theorem}[Correctness of Transformation] Any transformed program is
semantically equivalent to its source: 
\label{thm:correctness}
	\begin{multline*}
	\forall \cmd, \cmdt, \Mem, \Mt, \Mem', \Mt', \actset, \actset'.\\
	\configTwo{\actset}{\cmd} \tto \configTwo{\actset'}{\cmdt} 
	\AND \configTwo{\Mem}{\cmd} \To^* \configTwo{\Mem'}{\Skip} \\
	\AND \configTwo{\Mt}{\cmdt} \To^* \configTwo{\Mt'}{\Skip}
	\AND  \Mem = \Mt^{\actset} \\
	\sat \Mem' = (\Mt')^{\actset'}.
	\end{multline*}
\end{Theorem} 
\begin{proofsketch}
By induction on the transformation rules. The full proof is available in
\ifreport
Appendix~\ref{appendix:correcttrans}.
\else
the full version of this paper~\cite{deptypefull}.
\fi
\end{proofsketch}

\subsection{Relation to Information Flow Analysis}
\label{sec:interface}
Up to this point, it might be unclear why introducing extra variables can
improve the precision of information flow analysis. We first note that
transformed programs enable more precise reasoning for dataflows. Consider the
program in Figure~\ref{fig:example:flow} and
Figure~\ref{fig:example:flow:trans}. In the transformed program, it is clear
that the value stored in $x$ never flows to variable $\low$; but such 
information
is not obvious in the source program. Moreover, Theorem~\ref{thm:correctness}
naturally enables a more precise analysis of the transformed program, since
it implies that \emph{if any property holds on the final active set $\actset'$
for the transformed program, then the property holds on the entire final memory
for the original program}. That is, in terms of information flow security, the
original program leaks no information if the transformed program leaks no
information in the \emph{subset} $\actset'$ of the final memory. Consider the
example in Figure~\ref{fig:example:flow:trans} again.
Theorem~\ref{thm:correctness} allows a program analysis to accept the (secure)
program even though the variable $x$, which is not in $\actset'$, may leak the
secret value.

In Section~\ref{sec:comparison}, we show that, in general, the program
transformation automatically makes a flow-insensitive type system (e.g., the
Volpano, Smith and Irvine's system~\cite{vsi96} and the system in
Section~\ref{sec:typesystem}) at least as precise as a classic
flow-sensitive type system~\cite{Hunt:flowsensitive}.

\subsection{Relation to Single Static Assignment (SSA)}
\label{sec:ssa}

SSA~\cite{ssa} is used in the compilation chain to improve and simplify
dataflow analysis. Viewed in this way, it is not surprising that our program
transformation shares some similarity with the standard SSA-transformation.
However, our transformation is different from the latter in major ways:

\begin{itemize}

\item Most importantly, our transformation does not involve the
distinguishing $\phi$-functions of SSA. First of all, removing $\phi$-functions
simplifies the soundness proof, since the resulting target language syntax and
semantics are completely standard. Moreover, it greatly
simplifies information flow analysis on the transformed programs. Intuitively,
the reason is that in the standard SSA from, the $\phi$-function is added after
a branch (i.e., in the form of $(\ifcmd{e}{c_1}{c_2});x:=\phi(x_1,x_2)$).
However, without a nontrivial program analysis for the $\phi$-function, the path
conditions under which $x:=x_1$ and $x:=x_2$ occur (needed for path-sensitivity)
is lost in the transformed program. On the other hand, extra assignments are
inserted under the corresponding branches in our transformation. The
consequence is that the path information is immediately available for the
analysis on the transformed program. We defer a more detailed discussion on
this topic to Section~\ref{sec:brackettotype}, after introducing our type
system.

\item As discussed in Section~\ref{sec:interface}, the final active set
$\actset'$ generated from the transformation is crucial for enabling a more
precise program analysis on the transformed program (intuitively, an
information flow analysis may safely ignore variables not in $\actset'$);
however, such information is lost in the standard SSA form.


\item Our general transformation offers a full spectrum of analysis precision:
from adding no active copy to adding one copy for each assignment, but the
standard SSA transformation only performs the latter.
\end{itemize}

\section{Type System}
\label{sec:typesystem}

The second component of the analysis is a sound type system with expressive 
dependent labels. The
type system analyzes a transformed program along with the final active set; the
type system ensures that the final values of the public variables in the final
active set are not influenced by the initial values of secret variables.

\subsection{Overview}
\label{sec:challenge}

We first introduce the nonstandard features in the type system: dependent
security labels and program predicates.

Return to the example in Figure~\ref{fig:example:path}. We observe that this
program is secure because:
\begin{inparaenum}[1)]
\item $y$ holds a secret value only when $ \low_1 < 0$, and
\item the information flow from $y$ to $x$ at line 3 only occurs when  $ \low_1 > 0$.
\end{inparaenum}
Accordingly, to gain path-sensitivity, two pieces of information are needed  in
the type system:
\begin{inparaenum}[1)]
\item expressive security labels that may depend on program states, and
\item an estimation of program states that may reach a program point.
\end{inparaenum}

We note that such information can be gained by introducing \emph{dependent
security labels} and \emph{program predicates} to the type system. For the
example in Figure~\ref{fig:example:path}, the relation between the level of $ y
$ and the value of $ x $ can be described as a concise dependent label
$(\low_1<0?\High:\Low)$, meaning that the security level of $x$ is $\High$ when
$\low_1<0$; the level is $\Low$ otherwise. Moreover, for precision, explicit
and implicit flows should only
be checked under program states that may reach the program point. In general, a
predicate overestimates such states. For the example in
Figure~\ref{fig:example:path}, checking that the
explicit flow from $y$ to $x$ is secure under any program state is too
conservative, since it only occurs when $\low_1>0 $. With a program predicate
that $\low_1>0$ for the assignment $x:=y$,
the label of $y$ can be precisely estimated as $\Low$.  Note that our analysis
agrees with the definition of path-sensitivity: it understands that the two
assignments $ y:=\high$ and $ x:=y;$ never execute together in one execution. The
example in Figure~\ref{fig:example:path} is accepted by our type system.


\subsection{Challenge: Statically Checking Implicit Declassification}
\label{sec:impdeclassify}

Though designing a dependent security type system may seem simple at the first
glance, handling mutable variables can be challenging. The implicit
declassification problem, as defined in \cite{hwtiming}, occurs whenever the
level of a variable changes to a less restrictive one, but its value remains
the same. Consider the insecure program in
Figure~\ref{fig:example:impDeclass:insecure}, which is identical to the secure
program in Figure~\ref{fig:example:path} except for line 4. This program is
obviously insecure since the sequence $y:=\high;\low_1:=1;x:=y;\low_2:=x;$ may be
executed together.  Compared with Figure~\ref{fig:example:path}, the root cause
of this program being insecure is that at line 4 (when $\low_1$ is updated),
$y$'s new level \Low (according to the label $\low_1<0?\High:\Low$) is no
longer consistent with the value it holds.

\lstset{numbers=left}
\newsavebox\impDeclassExample
\begin{lrbox}{\impDeclassExample}
\ifreport
\begin{minipage}{0.3\textwidth}
\else
\begin{minipage}{0.22\textwidth}
\fi
\begin{lstlisting}
$x:=0$; $y:=0$;
$\If$ ($\low_1$ < 0) $\Then$ 
	$\assign{y}{\high}$; 
$\bkassign{\low_1}{1}$;
$\If$ ($\low_1$ > 0) $\Then$
	$\assign{x}{y}$; 
$\assign{\low_2}{x}$; 
\end{lstlisting}
\end{minipage}
\end{lrbox}
\newsavebox\impDeclassExampleTrans
\begin{lrbox}{\impDeclassExampleTrans}
\ifreport
\begin{minipage}{0.3\textwidth}
\else
\begin{minipage}{0.22\textwidth}
\fi
\begin{lstlisting}
$x:=0$; $y:=0$;
$\If$ ($\low_1$ < 0) $\Then$
	$\assign{y}{\high}$; 
$\assign{\low_3}{1}$;
$\If$ ($\low_3$ > 0) $\Then$
	$\assign{x}{y}$; 
$\assign{\low_2}{x}$; 
\end{lstlisting}
\end{minipage}
\end{lrbox}

\begin{figure}
	\centering
	\subfigure[Insecure Program. ]{
		\label{fig:example:impDeclass:insecure}\
		\usebox\impDeclassExample
	}
	\subfigure[Transformed Program of \ref{fig:example:impDeclass:insecure}. ]{
		\label{fig:example:impDeclass:trans}\
		\usebox\impDeclassExampleTrans
	}
	\caption{Examples: Implicit Declassification.}
	\label{fig:example:impDeclass}
\end{figure}

The type systems in~\cite{hwtiming,trustzone} resort to a run-time mechanism to tackle
the implicit declassification problem. However, that also means that the type system
might change the semantics of the program being analyzed. In this paper, we aim
for a purely static solution.

\paragraph{Program Transformation and Implicit Declassification}
Although the program transformation in Section~\ref{sec:transformation} is
mainly designed for flow-sensitivity, we observe that it also helps to detect implicit
declassification. Consider the example in
Figure~\ref{fig:example:impDeclass:insecure} again, where the assignment at
line 4 has brackets. The corresponding transformed program
(Figure~\ref{fig:example:impDeclass:trans}) does not have an implicit
declassification problem since updating $\low_3$ at line 4 does not change
$y$'s level, which depends on the value of $\low_1$, rather than $\low_3$.
Moreover, the insecure program cannot be type-checked since both ``then''
branches might be executed together.

While adding extra variable copies helps in the previous example, it unfortunately
does not eliminate the issue. The intuition is that even for a fully-bracketed
program, variables modified in a loop might still be mutable (since the local
variables defined in the loop might change in each iteration).
Consider the program in \ref{fig:example:loopDeclass:insecure}. This program is
insecure since it copies $\high$ to $y$ in the first iteration, and copies
$y$ to $\low$ in the next iteration. When fully-bracketed, the loop body
becomes
\begin{lstlisting}[numbers=none,xleftmargin=5ex]
$\ifcmd{x_2\%2=0}{y_1:=\high;y_3:=y_1}{\dots}$;
$x_3:=x_2+1;x_2:=x_3; y_2:=y_3$;
\end{lstlisting}
where the labels of $y_1$ and $y_3$ depend on $x_2$. In this program, implicit
declassification happens when $x_2$ is updated.

\lstset{numbers=left}
\newsavebox\DeclassLoopSecure
\begin{lrbox}{\DeclassLoopSecure}
\ifreport
	\begin{minipage}{0.3\textwidth}
\else
	\begin{minipage}{0.22\textwidth}
\fi
\begin{lstlisting}
$x=0$;
$\While$ ($x<10$) {
  $\If$ ($x\%2\!=\!0$) $\Then$
    $\assign{y}{\high}$;
  $\Else$ 
    $\low:=y$;
  $\assign{x}{x+1}$; 
}
\end{lstlisting}
	\end{minipage}
\end{lrbox}

\newsavebox\DeclassLoopInsecure
\begin{lrbox}{\DeclassLoopInsecure}
	\begin{minipage}{0.2\textwidth}
\begin{lstlisting}
$x=0$;
$\While$ ($x<10$) {
  $\If$ ($x\%2\!=\!0$) $\Then$
    $\assign{y}{\high}$;
  $\Else$
    $\low:=y$;
  $\assign{x}{x+1}$; 
  $\assign{y}{0}$; 	
}
\end{lstlisting}
	\end{minipage}
\end{lrbox}
\begin{figure}
	\centering
	\subfigure[Insecure Program. ]{
		\label{fig:example:loopDeclass:insecure}\
		\usebox\DeclassLoopSecure
	}
	\subfigure[Secure Program.]{
		\label{fig:example:loopDeclass:secure}\
		\usebox\DeclassLoopInsecure
	}
	\caption{Examples: Implicit Declassification in Loop.}
	\label{fig:example:loopDeclass}
\end{figure}

One naive solution is to disallow mutable variables in a program. However,
dependence on mutable variables does not necessarily break security.
Consider the program in Figure~\ref{fig:example:loopDeclass:secure}, which is
identical to the previous example except that $y$ is updated at line 8. In this
program, $y$'s level depends on the mutable variable $x$, but it is secure
since the value of $\high$ never flows to the next iteration.

\paragraph{Our Solution}
Our insight is that changing $y$'s level at line 7 in
Figure~\ref{fig:example:loopDeclass:secure} is secure since the value of $y$ is
not used in the future (in terms of dataflow analysis, $y$ is dead after line
6). This observation motivates us to incorporate a customized \emph{liveness
analysis} (Section~\ref{sec:liveness}) into the type system: an update to a
variable $x$ is allowed if no labels of the \emph{live variables} at that
program point depend on $x$.

\subsection{Type Syntax and Typing Environment}
\begin{figure}
\begin{align*}
	&\text{Level}\quad  &\ell & ∈     \Lattice \\[-1ex]
	&\text{Label}\quad  &\Type & ::=   \ell \mid e?\tau_1:\tau_2 \mid \Type_1 \join \Type_2 \mid \Type_1 \meet \Type_2 
\end{align*}
\caption{Syntax of Security Labels.}
\label{fig:typing:syntax}
\end{figure}

In our type system, types are extended with security labels, whose syntax is shown in
Figure~\ref{fig:typing:syntax}.  The simplest form of label $\Type$ is a
concrete security level $\ell$ drawn from a security lattice $\Lattice$.
Dependent labels, specifying levels that depend on run-time values, have the form of 
$(e?\tau_1:\tau_2) $, where $ e $ is an expression. Semantically, if $ e $
evaluates to a non-zero value, the dependent label evaluates to $ \tau_1$, 
otherwise, 
$\tau_2$. A security label can also be the least upper bound, or
the greatest lower bound of two labels.

We use $\G$ to denote a typing environment, a function from program variables
to security labels. The integration of dependent labels puts constraints on the
typing environment $\G$ to ensure soundness. In particular, we say $ \G $ is
\emph{well-formed}, denoted as $ \proves \G $, if:
\begin{inparaenum}[1)]
\item no variable depends on a more restrictive variable, preventing 
leakage from labels;
\item there is no chain of dependency.
\end{inparaenum} 
These restrictions are formalized as follows, where $\free(\tau)$ denotes the
free variables in $ \tau $:
\begin{Definition}[Well-Formedness] A typing environment $\G$ is well-formed,
written $\proves \G$, if and only if:
\begin{align*}
\forall x \in \Vars.~~ & (\forall x' \in \free(\G(x)).~\G(x') \absleq \G(x))\\
	 \AND & (\forall x' \in \free(\G(x)).~\free(\G(x')) = \emptyset )
\end{align*}
\end{Definition}

We note that the definition rules out self-dependence, since if $x\!\in\!
\free(\G(x))$, we have $\free(\G(x))\!=\!∅$. Contradiction.

\subsection{Predicates and Variable Liveness}

Our type system is parameterized on two static program analyses: a predicate
generator and a customized liveness analysis. Instead of embedding these
analyses into our type system, we follow the modular design introduced in
\cite{hwtiming} to decouple program analyses from the type system.
Consequently, the soundness of the type system is only based on the correctness
of those analyses, regardless of the efficiency or the precision of those analyses. 

\paragraph{Predicate Generator} 

We assume a predicate generator that generates a (conservative) program
predicate for each assignment $η$ in the transformed program, denoted as $
\hypo(η) $. A predicate generator is correct as long as each predicate is
always true when the corresponding assignment is executed.

A variety of techniques, regarding the trade-offs between precision and
complexity, can be used to generate predicates that describe the run-time
state. For example,
weakest preconditions \cite{Dijkstra75} or the linear propagation
\cite{hwtiming} could be used. Our observation is that for path-sensitivity,
only shallow knowledge containing branch conditions is good enough for our type
system.  

\paragraph{Liveness Analysis} 
\label{sec:liveness}
Traditionally, a variable is defined as alive if its value will be read in the
future. But in our type system, if a variable $x$ is alive, then any free
variable in the label of $x$ should also be considered as alive, because the
concrete level of $x$ depends on those variables. Moreover, we assume at the
end of a program, only the variables in the final active set are alive,
due to Theorem~\ref{thm:correctness}.

\begin{figure}
	\begin{align*}
	\liveout{\text{final}} =&~ \actset\\
	\livein{s} = &~ \gen{s} \cup (\liveout{s} - \kil{s})\\
	\liveout{s} = &~ \bigcup _{p\in succ[s]}\livein{p}\\
	\gen{\assigneta{x}{e}} =&~  \free(e) \cup (\bigcup_{v \in  \free(e)} 
	\free(\Gamma(v)))  \\
	\kil{\assigneta{x}{e}} =&~ \{x\} 
	\end{align*}
	\caption{Liveness Analysis of $ \liveset{\actset} $.}
	\label{fig:live}
\end{figure}
The liveness analysis is defined in Figure~\ref{fig:live}, where $s$
denotes a program command, and \textit{final} refers to the last command of the
program being analyzed. Here, \textit{final} is the initial state for the
backward dataflow analysis. $succ[s]$ returns the successors (as a set) of
the command $s$. In the GEN set of
an assignment $\assign{x}{e}$, both $ \free(e) $, and $ \bigcup_{v \in
\free(e)} \free(\G(v))$, the free variables inside their labels, are included.
Since we are analyzing the transformed program, the state of the final active set
is crucial for precision. Therefore, the analysis also enforces that, at the end of the
program, all active copies in $ \actset $ are alive. Other rules are
standard for liveness analysis.

\paragraph{Interface to the Type System}

We assume each assignment in the transformed language is associated with a
unique identifier $ \eta $. We use $ \before \eta $ and $ \after \eta $ to
denote the precise program points right before and after the assignment
respectively. For example, $ \hypo(\before \eta) $ represents the predicates
right before statement $ \eta $, and $ \liveset{\actset}(\after \eta) $ denotes
the alive set right after statement $ \eta $ with initialization of $ \actset $
as the final live set.

\subsection{Typing Rules}
\label{sec:typingrules}
\begin{figure*}
\begin{mathpar}
\inferrule*[right=T-Const]{ }{ \G  \proves n : \bot }
\and
\inferrule*[right=T-Var]{	\G(x) = \Type}
     {\G  \proves x : \Type}
\and
\inferrule*[right=T-Op]{\G \proves e : \Type_1 \and \G \proves e' : \Type_2}
     {\G \proves e\ \op\ e' : \Type_1\join \Type_2}
 \end{mathpar}
\caption{Typing Rules: Expressions.}
\label{fig:while-typing:exp}
\begin{mathpar}
\inferrule*[right=T-Skip]
{ }{\G, \pc \proves \Skip   }
\and
\inferrule*[right=T-Seq]
{\G, \pc \proves \cmd_1 
	\quad \G, \pc \proves \cmd_2  }
{\G, \pc \proves \cmd_1;\cmd_2  }
\and
\inferrule*[right=T-If]
{\G \proves e:\Type 
	\quad \G, \Type \join \pc \proves \cmd_1 
	\quad \G, \Type \join \pc \proves \cmd_2  }
{\G, \pc \proves \ifcmd{e}{\cmd_1}{\cmd_2}   }
\and
\inferrule*[right=T-Assign]
{\G \proves e: \Type 
	\quad \valid{\hypo(\before{\eta}) \sat \Type \join \pc \absleq \G(x)} 
	\quad { \forall v \in \liveset{\actset}(\after \eta).  x \not \in \free(\G(v))}	 }
{ \G, \pc \proves \assigneta{x}{e}   }
\and
\inferrule*[right=T-While]
{\G \proves e:\Type 
	\quad
	\G, \Type \join \pc \proves \cmd  }
{\G, \pc \proves \while{e}{\cmd} }

\end{mathpar}
\caption{Typing Rules: Commands.}
\label{fig:while-typing:commands}

\end{figure*}

The type system is formalized in Figure~\ref{fig:while-typing:exp} and
Figure~\ref{fig:while-typing:commands}.  Typing rules for expressions have the
form of $\G \proves \expr : \Type$, where $\expr$ is the expression being
checked and $\Type$ is the label of $e$. The typing judgment of commands has
the form of $\G, \pc \proves \cmd$. Here, $\pc$ is the usual program-counter
label~\cite{sm-jsac}, used to control implicit flows. 

Most rules are standard, thanks to the modular design of our type system. The
only interesting one is rule~\ruleref{T-Assign}. For an assignment $
\assigneta{x}{e}$, this rule checks that both the explicit and implicit flows
are allowed in the security lattice: $\Type \join \pc \absleq \G(x)$. Note that
since $\tau$ might be a dependent label that involves free program variables,
the $\absleq$ relation is technically the lifted version of the relation on the
security lattice. Hence, the constraint $\Type \join \pc \absleq \G(x)$
requires the label of $ x $ to be at least as restrictive as the label of current
context $ \pc $ and the label $e$ under any program execution. For precision,
the type system validates the partial ordering under the predicate
$\hypo(\before\eta)$, the predicate that must hold for any execution that
reaches the assignment.

Moreover, the assignment rule checks that for any variable in the liveness set
after the assignment, its security label must not depend on $x$; otherwise, its
label might be inconsistent with its value. As discussed in
Section~\ref{sec:impdeclassify}, this check is required to rule out insecure
implicit declassification.

At the top level, the type system collects proof obligations in the form of
$ \valid{\hypo \sat \tau_1 \absleq \tau_2} $, where $ \tau_1 $ and $ \tau_2 $
are security labels, and $\hypo$ is a predicate. Such proof obligations can
easily be discharged by theorem solvers, such as Z3~\cite{z3}.

As an example, consider again the interesting examples in
Figure~\ref{fig:example:loopDeclass}. In both programs, we can assign $y$ to
the dependent label $ (x \% 2=0?\High:\Low) $, and assign $x$ to the label
$\Low$. From the liveness analysis, we know that the live sets right after line
7 are $ \{x,y, \high \} $ and $ \{x, \low, \high \} $ for
Figure~\ref{fig:example:loopDeclass:insecure} and
Figure~\ref{fig:example:loopDeclass:secure} respectively. Hence, the type
system correctly rejects the insecure program in
Figure~\ref{fig:example:loopDeclass:insecure} since the check at line 7,
$\forall v \in \liveset{\actset}(\after \eta).~x \not \in \free(\G(v))$, fails.
On the other hand, the check at line 7 succeeds for the program in
Figure~\ref{fig:example:loopDeclass:secure}. For line 4 in
Figure~\ref{fig:example:loopDeclass:secure}, the assignment rule generates one
proof obligation
\[ \valid{(x\%2=0) \sat \Low\join \High \absleq (x \% 2=0?\High:\Low)}\]
which is clearly true for any value of $x$. In fact, the secure program in
Figure~\ref{fig:example:loopDeclass:secure} is correctly accepted by the type
system in Figure~\ref{fig:while-typing:exp} and
Figure~\ref{fig:while-typing:commands}.


\subsection{Program Transformation and Information Flow Analysis}
\label{sec:brackettotype}

We now discuss the benefits of the program transformation in
Section~\ref{sec:transformation} for information flow analysis in details.

\subsubsection{Simplifying Information Flow Analysis}
As discussed in Section~\ref{sec:ssa}, our transformation does not involve the
distinguishing $\phi$-functions of SSA. Doing so simplifies information flow
analysis on the transformed programs.  We illustrate this using the following
example, where $y$ is expected to have the label $(x=1?\Low:\High)$ afterwards.
\begin{lstlisting}[numbers=none,xleftmargin=5ex]
$\ifcmd{x=1}{y:=0}{y:=\high}$
\end{lstlisting}
Our transformation yields the following program, which can be verified with
labels $y_1:\Low$, $y_2:\High$, $y_3: (x=1?\Low:\High)$.
\begin{lstlisting}[numbers=none,xleftmargin=5ex]
$\If~(x=1)~\Then~(y_1:=0;y_3:=y_1)$
$    \Else~{(y_2:=\high;y_3:=y_2)};$
\end{lstlisting}
In comparison, the standard SSA form is:
\begin{lstlisting}[numbers=none]
$(\ifcmd{x=1}{y_1\!:=\!0}{y_2\!:=\!\high};)y_3\!:=\!\phi(y_1,y_2);$
\end{lstlisting}
To verify this program, a type system would need at least a nontrivial typing
rule for $\phi$, which somehow ``remembers'' that $y_3 := y_2$ occurs only when
$x=1$. Even with such knowledge, the type of $y_2$  cannot simply be $\High$,
since otherwise, assigning $y_2$ to $y_3$ at $\phi$ is insecure. In fact, the
labels required for verification are $y_1, y_2, y_3: (x=1?\High:\Low)$.

Similar complexity is also involved for the $\phi$-functions inserted for
loops: to precisely reason about information flow, the semantics and typing
rules of $\phi$ also need to track the number of iterations.

\subsubsection{Improving Analysis Precision}
Precision-Wise, bracketed assignments improve analysis 
precision in two ways.
First, as discussed in Section~\ref{sec:interface}, they improve
flow-sensitivity by introducing new variable definitions. Second, they also
improve path-sensitivity by enabling more accurate program predicates. Consider
the following example. 
\begin{lstlisting}[numbers=none,xleftmargin=5ex]
$x:=-1;$ 
$\If~(x>0)~\Then~y:=\High;~\Else~ y:=1;$
$\bkassign{x}{-x};$
$\If~(x>0)~\Then~\low:=y;$ 
\end{lstlisting}
This program is secure since $\low$ becomes $1$ regardless of the value of
$\high$. However, without the bracket shown, the type system rejects it since no such
label $\tau_y$ satisfies the constraints that $(x>0) \Rightarrow (\High \absleq
\tau_y) $ (arising from the first if) and $(x>0) \Rightarrow (\tau_y \absleq \Low)
$ (arising from the second if).

However, with the bracket, the last two lines become
\begin{lstlisting}[numbers=none,xleftmargin=5ex]
$\assign{x_1}{-x};$
$\If~(x_1>0)~\Then~\low:=y; $
\end{lstlisting}
This program can be type-checked with y's label as $(x>0?\High:\Low)$ and a
precise enough predicate generator, which generates
$x_1=-x$ after the assignment $\assign{x_1}{-x}$, because constraints $(x>0)
\Rightarrow (\High \absleq \tau_y) $ and $(x_1>0∧ x_1=-x) \Rightarrow (\tau_y
\absleq \Low) $ can be solved with y's label mentioned above.
%

\section{Soundness}
\label{sec:soundness}

Central to our analysis is rigorous enforcement of a strong
information security property. We formalize this property in this section and
sketch a soundness proof. The complete proof is available in
\ifreport
Appendix~\ref{appendix:soundness}.
\else
the full version of this paper~\cite{deptypefull}.
\fi

\subsection{Noninterference}
\label{noninterference}

Our formal definition of information flow security is based on
noninterference~\cite{noninterference}. Informally, a program satisfies
noninterference if an attacker cannot observe any difference between two
program executions that only differ in their confidential inputs. This
intuition can be naturally expressed by \emph{semantics models} of program
executions.

Since a security label may contain program variables, its concrete level cannot
be determined statically in general. But it can always be evaluated under a
concrete memory:

\begin{Definition}
For a security label $\tau$, we evaluate its
concrete level under memory $\Mem$ as follows:
\[\valueof{\tau}{\Mem} = \lbl \text{, where } \configTwo{\Mem}{\tau} \evalto \lbl\]
\end{Definition}

Moreover, to simplify notation, we use $\glabelof{x}{\Mem}{\G}$ to denote the
concrete level of $x$ under $m$ and $\G$ (i.e., $\glabelof{x}{\Mem}{\G} =
\valueof{\G(x)}{\Mem}$).

To formally define noninterference in the presence of dependent labels, we first
introduce an equivalence relation on memories. Intuitively, two memories are
$(\G,\ell)$-equivalent if all variables with a level below level $\ell$ agree
on both their concrete levels and values.

\begin{Definition}[$(\G,\ell)$-Equivalence]
Given any concrete level $\lbl$ and $\G$, we say two memories $\Mem_1$ and
$\Mem_2$ are equivalent up to $\lbl$ under $\G$ (denoted by $\Mem_1
\lbleq{Γ}\Mem_2$) iff
\begin{multline*}
\forall x\in \Vars.\\[-.5ex]
(\glabelof{x}{\Mem_1}{\G} \LEQ \lbl \iff \glabelof{x}{\Mem_2}{\G} \sqsubseteq \lbl) 
\AND \glabelof{x}{\Mem_1}{\G} \LEQ \lbl \\
\implies \Mem_1(x) = \Mem_2(x)
\end{multline*}
\end{Definition}

It is straightforward to check that $\lbleq{\G}$ is an equivalence relation on
memories. Note that we require type of $x$ be bounded by $\lbl$ in $\Mem_2$
whenever $\glabelof{x}{\Mem_1}{\G} \LEQ \lbl$. The reason is to avoid label
channels, where confidential data is leaked via the security level of a
variable~\cite{Russo:CSF10,hwtiming}.

Given initial labels $\G$ on variables and final labels $\G'$ on variables, we
can formalize noninterference as follows:

\begin{Definition}[Noninterference]
We say a program $\cmd$ satisfies noninterference w.r.t. $\G$, $\G'$ if
equivalent initial memories produce equivalent final memories:
\begin{multline*}
\forall \Mem_1, \Mem_2, \lbl.\\[-.5ex]
 m_1\lbleq{\G} \Mem_2 \AND
\configTwo{\Mem_1}{c} \To^* \configTwo{\Mem_1'}{\Skip} \AND \\
\configTwo{\Mem_2}{c} \To^* \configTwo{\Mem_2'}{\Skip} \\ \implies \Mem_1'
\lbleq{Γ'} \Mem_2'
\end{multline*}
\end{Definition}

The main theorem of this paper is the soundness of our analysis: informally, if
the transformed program type-checks, then the original program satisfies
noninterference. Since the type system applies to the transformed program, we
first need to connect the types in the original and the transformed programs.
To do that, we define the projection of types for the transformed program in a
way similar to Definition~\ref{def:memactset}:

\begin{Definition}[Projection of Types]
\label{def:typeactset}
Given an active set $\actset$ and $\Gt$, types of variables in the transformed
program, we use $\Gt^{\actset}$ to denote a mapping from $\Vars$ to $\tau$ as
follows:
\[
∀v\in \Vars.~\Gt^{\actset}(v)=\Gt(\actset(v))
\]
\end{Definition}

Formally, the soundness theorem states that if a program $c$ under active set
$\actset$ (e.g., an identity function) is transformed to $\cmdt$ and final active set
$\actset'$, and $\cmdt$ is well-typed under the type system (parameterized on
$\actset'$), then $\cmd$ satisfies noninterference w.r.t. $\G^{\actset}$ and
$\G^{\actset'}$:

\begin{Theorem}[Soundness]
\label{thm:soundness}
\begin{multline*}
	\forall \cmd, \cmdt, \Mem_1, \Mem_2, \Mem_1', \Mem_2', \lbl, \Gt, \actset, \actset'~. \\
	\configTwo{\actset}{\cmd}\tto \configTwo{\actset'}{\cmdt} ∧\proves \Gt ∧ \Gt \proves \cmdt ∧
        \Mem_1\lbleq{\Gt^{\actset}} \Mem_2 ∧ \\
	\configTwo{\Mem_1}{c} \To^* \configTwo{\Mem_1'}{\Skip} \AND
	\configTwo{\Mem_2}{c} \To^* \configTwo{\Mem_2'}{\Skip} \\
	\implies \Mem_1' \lbleq{\Gt^{\actset'}} \Mem_2'
\end{multline*}
\end{Theorem}

\begin{figure}
    \centering
\ifreport
    \includegraphics[width=.55\columnwidth]{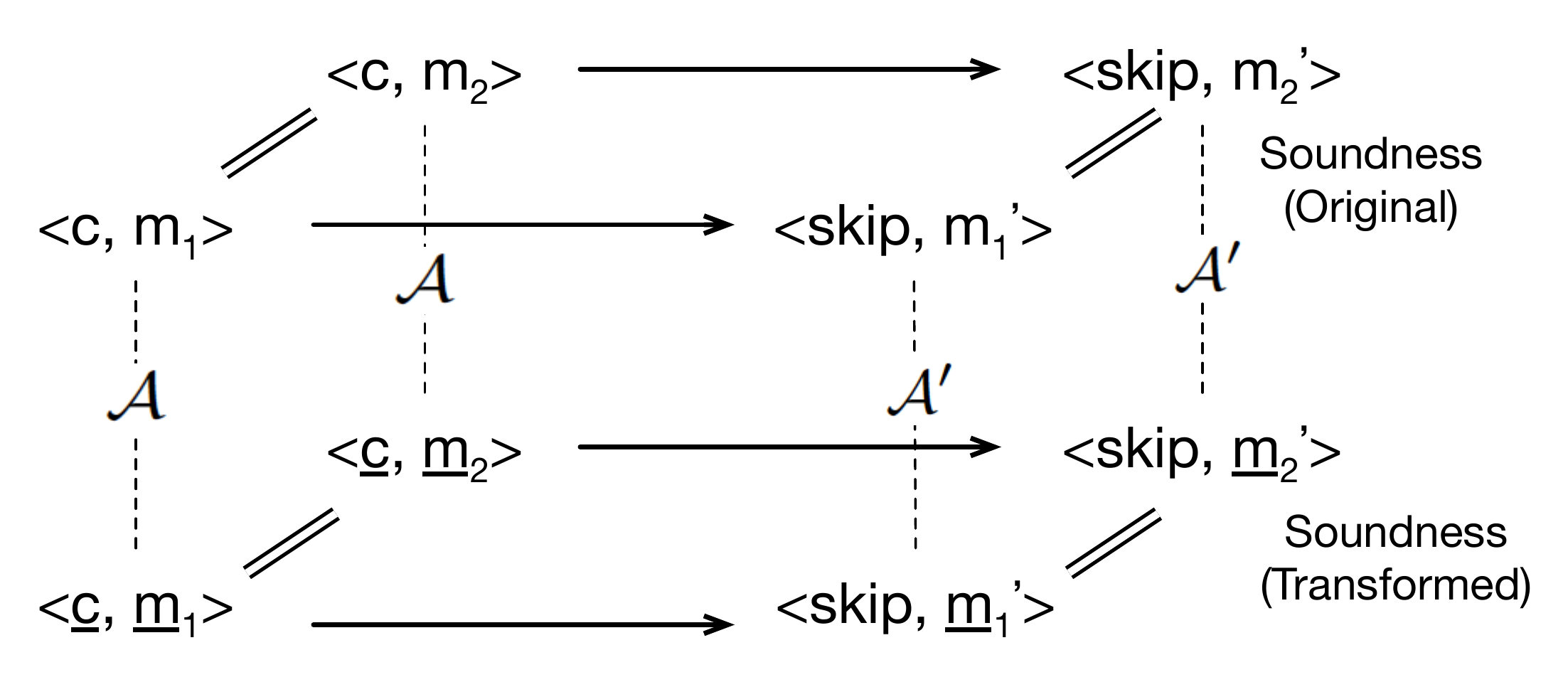}
\else
    \includegraphics[width=.9\columnwidth]{overview}
\fi
    \caption{Soundness of original and transformed programs.}
    \label{fig:soundness}
\end{figure}

To approach a formal proof, we notice that by the correctness of the program
transformation (Theorem~\ref{thm:correctness}), it is sufficient to show that
the transformed program leaks no information on the subset $\actset'$. Such
connection is illustrated in Figure~\ref{fig:soundness}. We formalize the
soundness for the transformed program w.r.t. initial and final active sets as
follows:

\begin{Theorem}[Soundness of Transformed Program]
 \label{thm:transsoundness}
 \begin{multline*}
\forall \cmdt, \Mt_1, \Mt_2, \Mt_3, \Mt_4, \lbl, \Gt, \actset, \actset'~. \\
         \configTwo{\actset}{\cmd}\tto \configTwo{\actset'}{\cmdt} ∧ \proves \Gt ∧
        \Gt \proves \cmdt~  \AND \Mt_1^{\actset} \lbleq{\Gt^{\actset}} \Mt_2^{\actset} \\
	\AND \configTwo{\Mt_1}{\cmdt} \To^* \configTwo{\Mt_3}{\Skip} \AND
 	\configTwo{\Mt_2}{\cmdt} \To^* \configTwo{\Mt_4}{\Skip} \\
	\implies \Mt_3^{\actset'} \lbleq{\Gt^{\actset'}} \Mt_4^{\actset'}
 \end{multline*}
 \end{Theorem}

\begin{proofsketch}
\begin{figure}
	\begin{mathpar}
		\switch(\Mt, x, \eta)(\varx) =
		\begin{cases}
			0, &\! x \in \free(\varx) ∧ x'\not\in \liveset{\actset'}(\eta)\\ 
			\Mt(\varx), &\! \text{otherwise}\\
		\end{cases}
	\end{mathpar}
	\[	
	\inferrule*[right=ST-Erase]
	{ \configTwo{\Mt}{x} \evalto n 
			\quad \Mt' = \Mt\Mupdate{x}{n}}
	{ \configTwo{\Mt}{\assigneta{x}{e}}\ERTo{\actset'} \configTwo{\switch(\Mt', x)}{\Skip}}	\\
	\]
	\caption{Erasure Semantics of Assignment.}
        \label{fig:erasure}	
\end{figure}


One challenge in the formal proof is that the equivalence relation
$\lbleq{\Gt}$ only holds on the active copies and it may break temporarily
during the program execution. Consider the example in
Figure~\ref{fig:example:loopDeclass:secure}. During the first iteration of the
loop body, $y$ holds a secret value but its level is $\Low$ right after line 8.
Hence, the relation $\lbleq{\Gt}$ may break at that point in the small-step
evaluation starting from two memories that only differ in secrets.
To tolerate such temporary violation of the $\lbleq{\Gt}$ relation, we prove
the soundness with a new semantics which \emph{enforces} that the relation
$\lbleq{\Gt}$ holds for all variables, and the final values of variables in
$\actset'$ agree with those in the standard semantics. The new semantics,
called the \emph{erasure semantics} is shown in Figure~\ref{fig:erasure}. The
semantics is parameterized on the final active set $\actset'$. The only
difference from the standard one is for assignments: the new assignment
rule~\ruleref{ST-Erase} sets variables that are not alive and whose
types depend on $x$ to be zero. It is easy to check that the erasure semantics
agrees on the final value of the variables in $\actset'$. Also, it removes the
temporary violation of the equivalence relation by forcing value of $y$ to be
zero after line 7 of Figure~\ref{fig:example:loopDeclass:secure}.  
The complete proof is available in
\ifreport
Appendix~\ref{appendix:soundness}.
\else
the full version of this paper~\cite{deptypefull}.
\fi
\end{proofsketch}

\section{Enabling Flow-Sensitivity with Program Transformation}
\label{sec:comparison}

Recall that the dependent security type system (without program transformation)
is flow-insensitive; yet, our program analysis is flow-sensitive with
the novel program transformation in Section~\ref{sec:transformation}.
In this section, we show that this is not a coincidence: the
program transformation automatically makes a flow-insensitive type system
(e.g., the Volpano, Smith and Irvine's system~\cite{vsi96} and the system in
Section~\ref{sec:typesystem}) flow-sensitive.

\subsection{The Hunt and Sands System}
\newcommand\fsproves{\proves_{\text{HS}}}
\begin{figure*}
\ifreport
\small
\else
\fi
\begin{mathpar}
\inferrule*[left=HS-Skip]
{ }{\pc \fsproves \G \{\Skip\} \G }
\and
\inferrule*[left=HS-Seq]
{\pc \fsproves \G \{\cmd_1\} \G'' 
	\quad \pc \fsproves \G'' \{\cmd_2\} \G'  }
{\pc \fsproves \G \{\cmd_1;\cmd_2\} \G'  }
\and
\inferrule*[left=HS-Assign]
{\G \fsproves \expr: \Type }
{ \pc \fsproves \G \{\assign{x}{e}\} \G\Mupdate{x}{\pc\join \Type}   }
\and
\inferrule*[left=HS-If]
{\G \fsproves e:\Type 
	\quad \Type \join \pc \fsproves \G \{\cmd_1\} \G_1 
	\quad \Type \join \pc \fsproves \G \{\cmd_2\} \G_2  
}
{\pc \fsproves \G \{\ifcmd{e}{\cmd_1}{\cmd_2}\} \G'  } 
\quad
\text{where } \G'=\G_1\join \G_2
\\
\inferrule*[left=HS-While]
{\G'_i \fsproves e:\Type_i
	\quad
	\Type_i \join \pc \fsproves \G'_i \{\cmd\} \G''_i
        \quad
        0\leq i \leq n  }
{\pc \fsproves \G \{\while{\expr}{\cmd}\} \G'_n }
\quad
\text{where } \G_0'=\G, \G'_{i+1}=\G''_i\join \G, \G'_{n+1}=\G'_n
\end{mathpar}
\caption{The Hunt and Sands System~\cite{Hunt:flowsensitive}.}
\label{fig:fstyping}

\end{figure*}

In~\cite{Hunt:flowsensitive}, Hunt and Sands define a classic flow-sensitive 
type
system where the security level of a program variable may ``float'' in the
program. In particular, Hunt and Sands (HS) judgments have the form of $\pc
\fsproves \G \{ \cmd \} \G'$, where $\G$ and $\G'$ are intuitively the typing
environments before and after executing $c$ respectively. 

Consider the program in Figure~\ref{fig:example:flow}. 
While a flow-insensitive type system rejects
it, the HS system accepts it with the following typing environments:
\[
\G \{\assign{x}{\high};\} \G \{\assign{x}{0};\} \G' \{\assign{\low}{x};\} \G'
\]
where $\G=\{x\mapsto \High, \low\mapsto \Low\}$ and $\G'=\{x\mapsto \Low,
\low\mapsto \Low\}$.

The HS typing rules for commands are summarized in Figure~\ref{fig:fstyping}.
We use $\fsproves$ to distinguish those judgments from the ones in our
system.  The interesting rules are rule~\ruleref{HS-If} and
rule~\ruleref{HS-While}: the former computes the type for each variable as the
least upper bound of its labels in the two branches; the latter computes the
least fixed-point of a monotone function (the while loop) on a finite lattice.

\subsection{Program Transformation and Flow-Sensitivity}

We show that the program transformation in Section~\ref{sec:transformation}
along with a flow-insensitive type system subsumes the HS system: for any
program $\cmd$ that can be type-checked in the HS system, the transformed
program of $\bkcmd{\cmd}$ (i.e., \emph{a fully-bracketed program}) can be type-checked
in a flow-insensitive type system. This result has at least two interesting
consequences:
\begin{enumerate}
\item The program transformation removes the source of ``flow-insensitivity'';
a flow-insensitivity type system can be automatically upgraded to a
flow-sensitive one.

\item The flow- and path-sensitive system in this paper strictly subsumes the
HS system: any secure program accepted by the latter is accepted by the former,
but not vise versa (e.g., the program in Figure~\ref{fig:example:path}).
\end{enumerate}

To construct types in the transformed program, we first introduce a few
notations. Given a typing environment $\G: \Vars\to \Type$ for the original
program and an active set $\actset$, we can straightforwardly construct a
(minimal) typing environment, written $\G_{\actset}$, whose projection on $\actset$ is $\G$: 
\[∀\vt\in \actset.~\G_{\actset}(\vt)\defn\G(\proj{\vt})\]  
Easy to check that $(\G_\actset)^{\actset}=\G$. 

Moreover, given a sequence of tying environments for the transformed program,
say $\Gt_1,\Gt_2,\dots$, we define a merge function, denoted as $\gmerge$, that
returns the union of $\Gt_1,\Gt_2,\dots$ so that conflicts in the environments
are resolved in the order of $\Gt_1,\Gt_2,\dots$. For example,
$\gmerge(\{x_1\mapsto \High, y_2\mapsto \Low\}, \{x_1\mapsto \Low, y_2\mapsto
\Low\})=\{x_1\mapsto \High, y_2\mapsto \Low\}$.

\begin{figure*}
\ifreport
\footnotesize
\fi
\begin{mathpar}
\inferrule*[left=C-Skip]
{ }{(\pc, \G,\actset) \{\Skip \tto \Skip\} (\G,\actset) \produce \G_{\actset} }
\and
\inferrule*[left=C-Seq]
{(\pc, \G, \actset) \{\cmd_1\tto \cmdt_1\} (\G'', \actset'')\produce \Gt_1
  \quad (\pc, \G'', \actset'') \{\cmd_2\tto \cmdt_2\} (\G', \actset') \produce \Gt_2}
{(\pc, \G, \actset) \{\cmd_1;\cmd_2\tto \cmdt_1;\cmdt_2\} (\G',\actset') \produce \gmerge(\Gt_1,\Gt_2) }
\and
\inferrule*[left=C-Assign]
{\pc \fsproves \G \{\assign{x}{e}\} \G\Mupdate{x}{\tau} 
  \quad \configTwo{\actset}{\bkassign{x}{e}}\tto 
  \configTwo{\actset\Mupdate{x}{x_i}}{\assign{x_i}{\et}}}
{ (\pc, \G, \actset) \{\bkassign{x}{e}\tto \assign{x_i}{\et}\} (\G\Mupdate{x}{\tau},
\actset\Mupdate{x}{x_i})\produce \G_{\actset}∪\{x_i\mapsto \tau\} }
\and
\inferrule*[left=C-If]
{   \inferrule{}{\G \fsproves e:\Type  \\\\ \configTwo{\actset}{e}\tto \et}
	\quad \inferrule{}{
        (\tau \join \pc, \G, \actset) \{\cmd_1\tto \cmdt_1\} (\G_1,\actset_1) \produce \Gt_1 \\\\
	(\tau \join \pc, \G, \actset) \{\cmd_2\tto \cmdt_2\} (\G_2,\actset_2) \produce \Gt_2}
        \quad \Phi(\actset_1, \actset_2)\tto \actset_3 \quad \G'=\G_1\join \G_2
}
{(\pc, \G, \actset) \{\ifcmd{e}{\cmd_1}{\cmd_2}\tto \ifcmd{\et}{(\cmdt_1;\assign{\actset_3}{\actset_1})}{(\cmdt_2;\assign{\actset_3}{\actset_2})}\} (\G',\actset_3) \produce \gmerge(\Gt_1,\Gt_2,\G'_{\actset_3}) } 
\\
\inferrule*[left=C-While]
{\inferrule{}{\pc \fsproves \G \{\while{\expr}{\cmd}\} \G' \\\\ \G' \fsproves e:\Type} 
    \quad \configTwo{\actset}{\cmd}\tto \configTwo{\actset_1}{\cmdt_1}
    \quad \inferrule{}{\Phi(\actset,\actset_1) \tto \actset_2 \\\\ \configTwo{\actset_2}{e}\tto \et}
    \quad  (\tau \join \pc, \G', \actset_2) \{\cmd\tto \cmdt\} (\G', \actset_3) \produce \Gt_0 
    }
{(\pc, \G, \actset) \{\while{\expr}{\cmd}\tto \assign{\actset_2}{\actset}; \while{\et}{(\cmdt;\assign{\actset_2}{\actset_3})}\} (\G',\actset_2) \produce \gmerge(\Gt_0,\G_{\actset})}
\end{mathpar}
\caption{Type Construction in Transformed Program.}
\label{fig:construction}

\end{figure*}

For a fully bracketed program $\bkcmd{\cmd}$, we can inductively define the
construction of $\Gt$ as inference rules in the form of 
\[(\pc, Γ,\actset) \{\bkcmd{\cmd}\tto \cmdt\} (Γ',\actset')\produce \Gt \]
where $\pc, \G, \cmd, \G'$ are consistent with the HS typing rules in the form of
$\pc \fsproves \G \{\cmd\} \G'$; $\actset, \bkcmd{\cmd}, \actset', \cmdt$ are
consistent with the program transformation rules in the form of
$\configTwo{\actset}{\bkcmd{\cmd}}\tto \configTwo{\actset'}{\cmdt}$. $\Gt$ is the
constructed typing environment that, as we show shortly in
Theorem~\ref{thm:constructcorrect}, satisfies $\Gt, \pc\proves \cmdt$. The
construction algorithm is formalized in Figure~\ref{fig:construction}.

Most parts of the rules are straightforward; they are simply constructed to be
consistent with the HS typing rules and the transformation rules in
Figure~\ref{fig:transformation}. The following lemma makes such
connections explicit.

\begin{Lemma} 
\label{lem:consistency}
\begin{multline*}
∀\pc, Γ, Γ', \actset, \actset', \cmd, \cmdt.~
\pc \fsproves \G \{\cmd\} \G' ∧ \configTwo{\actset}{\bkcmd{\cmd}}\tto \configTwo{\actset'}{\cmdt}\\
⇒ ∃\Gt.~(\pc, Γ,\actset) \{\bkcmd{\cmd}\tto \cmdt\} (Γ',\actset')\produce \Gt
\end{multline*}
\end{Lemma}
\begin{proof}
By induction on the structure of $\cmd$. 
\end{proof}

To construct types for the transformed program: for $\Skip$, we use
$\G_{\actset}$ (the typing environment before this command); for assignment,
since $x_i$ must be fresh, we can
simply augment  $\G_{\actset}$ with $\{x_i\mapsto \tau\}$. Other rules simply
merge constructed types from subexpressions in a conflict-solving manner, using
$\gmerge$. An eagle-eyed reader may find the construction is intuitively
correct if \emph{there is no conflict at all} in the merge operations.

We show that there is no conflict during construction by two observations.
First, if a variable has the same active copy before and after transforming a
fully-bracketed command $\bkcmd{c}$, then its type must remain the same
(before and after $c$) in the HS system. This property is formalized as
follows: 

\begin{Lemma}
\label{lem:extension}
\begin{multline*}
\pc \fsproves \G \{\cmd\} \G' ∧ \configTwo{\actset}{\bkcmd{\cmd}}\tto
\configTwo{\actset'}{\cmdt} ⇒ \\ 
∀v\in \Vars.~(\actset(v)=\actset'(v))⇒(\G(v)=\G'(v))
\end{multline*}
\end{Lemma}
\begin{proofsketch}
By induction on the structure of $\cmd$. The most interesting cases are for
branch and loop.

\begin{itemize}
\item $\ifcmd{e}{c_1}{c_2}$: by the HS typing rule, $\pc \fsproves \G \{\cmd_1\}
\G_1 ∧ \pc \fsproves \G \{\cmd_2\} \G_2 ∧ \G'=\G_1\join\G_2$. By the transformation
rules, $\configTwo{\actset}{\bkcmd{\cmd_i}}\tto \configTwo{\actset_i}{\cmdt_i},
i\in \{1,2\}$. Suppose $\actset(v)\not=\actset_1(v)$, $\actset_1(v)$ must be a
fresh variable generated in $\cmdt_1$, and hence, cannot be in $\actset_2$.  By
the definition of $\Phi$, $\actset_3(v)$ must be fresh. This
contradicts the assumption $\actset(v)=\actset'(v)$. Hence, $\G(v)=\G_1(v)$ by
the induction hypothesis.  Similarly, we can infer that $\G(v)=\G_2(v)$. So
$\G'(v)=\G_1(v)\join\G_2(v)=\G(v)$.

\item $\while{\expr}{\cmd}$: By rule~\ruleref{TRSF-While}, we have
$\configTwo{\actset}{\bkcmd{\cmd}}\tto \configTwo{\actset_1}{\cmdt_1}$,
where $\actset'$ is $\actset_1$ in this case. Hence, by the assumption, 
we have 
$\actset(v)=\actset_1(v)$.  By rule~\ruleref{HS-While}, there is a sequence of
environments $\G'_{i}, \G''_{i}$ such that $\pc\join \tau_i \proves \G'_i \{c\}
\G''_i$. By the induction hypothesis, $\G''_i(v)=\G'_i(v)$. Since
$\G'_0=\G$ and $\G'_{i+1}=\G\join \G''_i$ in rule~\ruleref{HS-While}, we can
further infer that $\G_{i+1}'(v)=\G_i''(v)$.  Hence, we have
$\G'(v)=\G_n(v)=\G_0(v)=\G(v)$.
\end{itemize}
\end{proofsketch}

Second, the constructed environment is \emph{minimal}, meaning that it
just specifies types for the variables in $\actset$ and the freshly generated
variables in $\cmdt$ (denoted as $\FVars(\cmdt)$). 

For technical reasons, we formalize this property along with the main
correctness theorem of the construction, stating that the transformed program
$\cmdt$ type-checks under the constructed environment $\Gt$.
Note that given any $\actset$, a fully bracketed command $\bkcmd{\cmd}$ always
transforms to some $\cmdt$
and $\actset'$. Hence, by Lemma~\ref{lem:consistency}, the following theorem is
sufficient to show that our program analysis is at least as precise as the HS
system:

\begin{Theorem}
\label{thm:constructcorrect}
\begin{multline*}
∀\cmd, \cmdt,\pc, \actset, \actset', \G, \G', \Gt.~
(\pc, \G, \actset) \{\bkcmd{\cmd}\tto \cmdt\} (\G', \actset') \produce \Gt \\
⇒ \Domain(\Gt)⊆\actset∪\FVars(\cmdt) ∧ \Gt, \pc \proves \cmdt 
\end{multline*}
\end{Theorem}
\begin{proof} Complete proof is available in
\ifreport
Appendix~\ref{appendix:construct}.
\else
the full version of this paper~\cite{deptypefull}.
\fi
\end{proof}

An interesting corollary of Theorem~\ref{thm:constructcorrect} is that the
transformed program can be type-checked under the classic fixed-level system
in~\cite{vsi96} as well.

\begin{Corollary}
\label{col:levels}
Theorem~\ref{thm:constructcorrect} also applies to the type system in
Figure~\ref{fig:while-typing:exp} and Figure~\ref{fig:while-typing:commands}
with the restriction that all labels are security levels (i.e., non-dependent
labels), which is identical to the system in~\cite{vsi96}.
\end{Corollary}
\begin{proof}
We note that the construction in Figure~\ref{fig:construction} only uses the
non-dependent part of our type system. Given all labels are security levels, it
is straightforward to check that our type system degenerates to the system
in~\cite{vsi96}.  
\end{proof}

Theorem~\ref{thm:constructcorrect} has a strong prerequisite that all
assignments in the original program are bracketed. We note that the result
remains true when such prerequisite is relaxed. Intuitively, a bracket is
unnecessary when the old and new definitions have the same security label.
Otherwise, a bracket is needed for flow-sensitivity. For example, to gain
flow-sensitivity, only the second assignment in Figure~\ref{fig:example:flow}
needs a bracket. The strong prerequisite is used in
Theorem~\ref{thm:constructcorrect} to make the result general (i.e.,
type-agnostic).

According to Corollary~\ref{col:levels}, the result that any secure program
accepted by the HS system is accepted by our analysis is true even if all
dependent security labels degenerate to simple security levels. On the other
hand, introducing dependent security labels makes our analysis strictly more
precise than the HS system. For example, the program in
Figure~\ref{fig:example:path} cannot be verified without dependent security
labels, but it can be type-checked with a label $y:(\low_1<0?\High:\Low)$.

\subsection{Comparison with the transformation in~\cite{Hunt:flowsensitive}}
Hunt and Sands show that if a program can be type-checked in the HS system,
then there is an equivalent program which can be type-checked by a fixed level
system~\cite{Hunt:flowsensitive}. However, their construction of the equivalent
program is type guided, meaning that the program transformation assumes that  
security
labels have already been obtained in the HS system, while our
program transformation (Figure~\ref{fig:transformation}) is general and
syntax-directed. An interesting application of our transformation is to test 
the typeability of the HS system without obtaining the types needed in 
the HS system in the first place.

It is noteworthy that our transformation is arguably simpler than
the HS transformation since our rule for loop has no fixed-point construction
while the latter has one. The reason is that compared with the HS
transformation, the goal of our transformation is easier to achieve: our
transformation improves analysis precision, while the HS transformation infers
the type for each variable in a program.  For example, consider a loop with
only one assignment $x := x + 1$, and $x$ is initially $\Low$. In the HS system,
the transformed program is $x_{\Low} := x_{\Low} + 1$, where $x_{\Low}$ is the
public version of variable $x$. On the other hand, our transformation generates
$x_1 := x_2;x_2 := x_1 + 1$. From the perspective of inferring labels,
introducing $x_1$ and $x_2$ might seem unnecessary since they must have the
same label according to the type system.  However, doing so might improve
analysis precision (e.g., the type system can specify the dependencies on $x_1$ 
and
$x_2$ separately with two copies of $x$).

\section{Related Work}
\label{sec:related}
We refer to~\cite{sm-jsac} for a comprehensive survey of static information
flow analysis. Here, we focus on the most relevant ones.

\paragraph{Dependent Labels and Information Flow Security} 
Dependent types have been widely studied and have been applied to practical
programming languages (e.g., \cite{Xi:POPL99, Xi:xanadu, myers-popl99,
Condit:deputy, Cayenne, Murray16}).  New challenges emerge for information flow analysis,
such as
precise, sound handling of information channels arising from label changes.  

For security type systems, the most related works are
SecVerilog~\cite{hwtiming,trustzone}, Louren\c{c}o and Caires~\cite{Caires15} and
Murray~et~al.~\cite{Murray16}. SecVerilog is a Verilog-like
language with dependent security labels for verifying timing-sensitive
noninterference in hardware designs. The type systems in~\cite{hwtiming,
trustzone} are not purely static: they remove implicit declassification by a
run-time enforcement that modifies program semantics. A recent extension to
SecVerilog~\cite{Ferraiuolo:dac} alleviates such limitation by
hardware-specific static reasoning. However, those type systems do not handle
loops (absent in hardware description languages), which gives rise to new
challenges for soundness.  Moreover, they are not flow-sensitive. The
recent work~\cite{Caires15} also allows the security type to depend on runtime
values. However, the system is flow-insensitive, and it does not have a
modular design that allows tunable precision. Moreover, the language has
limited expressiveness: it has no support for recursion, and it disallows
dependence on mutable variables. Exploring dependent labels to their full
extent exposes new challenges that we tackle in this work, such as implicit
declassification. 
Murray et al.~\cite{Murray16} present a flow-sensitive dependent security type
system for shared-memory programs. The type system enforces a stronger security
property: timing-sensitive non-interference for concurrent programs. However,
even when the extra complexity due to concurrency and timing sensitivity are
factored out, extra precision in their system is achieved via a floating
type system that tracks the typing environments and program states throughout
the program. In comparison, our analysis achieves flow-sensitivity via a
separate program transformation, which results in an arguably simpler type
system. Moreover, for dependency on mutable variables, their system only allows
a variable's security level to upgrade to a higher one, while our system allows a
downgrade to a lower level when doing so is secure.

Some prior type systems for information flow also support limited forms of
dependent
labels~\cite{myers-popl99,zm07,tse2007run,Jia:aura,Swamy:fable,Grabowski09}.
The dependence on run-time program state, though, is absent in most of these,
and most of them are flow- and path-insensitive.

\paragraph{Flow-Sensitive Information Flow Analysis}

Flow-sensitive information flow control~\cite{Russo:CSF10, Hunt:flowsensitive,
Austin:partialleak} allows security labels to change over the course of
computation. Those systems rely on a floating type system or
a run-time monitor to track the security labels at each program point. On the
other hand, the program transformation in our paper eliminates analysis
imprecision due to flow-insensitivity.  
Moreover, the bracketed assignments in a source program provide
tunable control for needed analysis precision. These features offer better
flexibility and make it possible to turn a flow-insensitive analysis to be
flow-sensitive.

\paragraph{Semantic-Based Information Flow Analysis}

Another direction of information flow security is to verify the semantic
definition of noninterference based on program logics. The first work that used
a Hoare-style semantics to reason about information flow is by Andrews and
Reitman~\cite{Andrews80}. Independence analysis based on customized
logics~\cite{Amtoft04,Amtoft06,Amtoft07} was proposed to check whether two
variables are independent or not. 
Self-Composition~\cite{Darvas05,selfcomposition} composes a program with a copy
of itself, where all variables are renamed. The insight is that noninterference
of a program $P$ can be reduced to a safety property for the self-composition
form of $P$.

Relational Hoare Logic~\cite{Benton04} was first introduced for a core imperative program to
reason about the relation of two program executions. It was
later extended to verify security proofs of cryptographic
constructions~\cite{Barthe09} and differential privacy of randomized
algorithms~\cite{Barthe12,Barthe15}. In the context of information flow
security, Relational Hoare Type Theory~\cite{rhtt} extends Hoare Type Theory
and has been used to reason about advanced information flow policies.

Though some semantic-based information flow analyses are flow- and
path-sensitive, most mechanisms incur heavy
annotation burden and steep learning curve on programmers. We believe our
approach shows that it is not necessary to resort to those heavyweight methods
to achieve both flow- and path-sensitivity.

\lstset{numbers=left}
\newsavebox\HighBranchUpgrade
\begin{lrbox}{\HighBranchUpgrade}
	\begin{minipage}{0.3\textwidth}
		\begin{lstlisting}[numbers=left,xleftmargin=15pt,framexleftmargin=15pt]
$x:=1; y:=1$;
$\If~(\high==0)~\Then~\Skip$
$\Else$ $x:=y$;
$\low := x$;
		\end{lstlisting}

	\end{minipage}
\end{lrbox}

\begin{figure}
	\centering
	\usebox\HighBranchUpgrade
	\caption{False Control-Flow Dependency.}
	\label{fig:example:highbranch}
\end{figure}

\section{Conclusions and Future Work}
\label{Sec:conclusion}

This paper presents a sound yet flow- and path-sensitive information flow 
analysis. The proposed analysis
consists of a novel program transformation as well as a dependent security
type system that rigorously controls information flow. We show that our
analysis is both flow- and path-sensitive.  Compared with existing work, we show
that our analysis is strictly more precise than a classic 
flow-sensitive
type system, and it tackles the tricky implicit declassification issue
completely at the compile time. Moreover, the novel design of our analysis allows a
user to control the analysis precision as desired. We believe our analysis offers
a lightweight approach to static information flow analysis along with 
improved precision.

The proposed analysis alleviates analysis imprecision due to
data- and path-sensitivity, but it still may suffer from other sources
of imprecision, such as the presence of insecure dead code and false
control-flow dependency. For example, consider the secure program in
Figure~\ref{fig:example:highbranch} (simplified from an example
in~\cite{besson13}) 
with security labels $\high:\High$, $\low:\Low$. In this example, although
$x$ is updated under a confidential branch condition, both branches result in
the same state where $x = 1$; thus, the outcome of $\low$ is independent of the 
value of $\high$. However,
our analysis rejects this program since rule~\ruleref{T-Assign}
conservatively assumes that any public variable modified in a confidential 
branch would
leak information. Motivated by the type system in~\cite{besson13}, a promising
direction that we plan to investigate is to incorporate sophisticated 
static program analyses so that the implicit flows can be ignored
for the variables whose values are independent of branch outcomes. 
Additionally, hybrid
information flow monitors (e.g.,~\cite{Russo:CSF10,besson13,Askarov15}) are
shown to be more precise than static flow-sensitive type systems. We plan to 
compare the analysis precision with those systems in our future work.

\section*{Acknowledgments}
We thank our shepherd Nataliia Bielova and anonymous reviewers for their
helpful suggestions. The noninterference proof in
\ifreport
Appendix~\ref{appendix:soundness}
\else
the full version of this paper~\cite{deptypefull}
\fi
is based on a note by Andrew Myers. This work was supported by NSF grant
CCF-1566411.

\balance
\def\IEEEbibitemsep{0pt plus 0pt} 
\bibliographystyle{IEEEtranS}
\bibliography{timing,master} 

\ifreport
\clearpage
\appendix

\section{Correctness of the Transformation}
\label{appendix:correcttrans}
We first show a few lemmas needed to prove the correctness of the program
transformation. 

\begin{Lemma}[Equal Expression] 
\label{lem:eqexp}
Any transformed expression $\et$ evaluates to the same value as in the original
program, and the transformation does not introduce fresh variables.
\begin{multline*}
	\forall e, \actset, \Mem, \Mt.\\
	\Mem = \Mt^{\actset} 
	\AND \configTwo{\actset}{e} \tto {\et}
	\AND \configTwo{\Mem}{e} \evalto n
	\AND \configTwo{\Mt}{\et} \evalto n'\\
	\sat n = n' ∧ \Vars(\et)⊆ \actset.
\end{multline*}
\end{Lemma}
\begin{proof} By induction on the structure of the expression $ e $:
	\begin{itemize}
		\item Case $ e = n $: trivial since $\et=n$ and $\Vars(\et)=∅$.
		\item Case $ e = x $: trivial since we have $\et=\actset(x)$ by
the transformation rule. By the assumption $\Mem=\Mt^{\actset}$, we have $\Mem(x) = \Mt^{\actset}(x) $. Moreover, $\Vars(\et)=\actset(x)\in \actset$.
	\item Case $ e = e_1 ~ \op ~ e_2 $: from the transformation, we know
that $\et$ has the form of $\et_1~ \op ~\et_2 $. By the induction hypothesis, we have
$ \configTwo{\Mem}{e_1} \evalto n_1$, $\configTwo{\Mem}{e_2} \evalto n_2$, $
\configTwo{\Mt}{\et_1} \evalto n_1'$, $\configTwo{\Mt}{\et_2} \evalto n_2' $,
and $n_1 = n_1' ∧ n_2 = n_2' $. Thus, we have  $e$ and $\et$ evaluate to the
same value. Moreover, $\Vars(\et)⊆\actset$ by the induction hypothesis.

	\end{itemize}
\end{proof}
\begin{Lemma}[Set-Assignment]
	\label{lem:phimem}
	\begin{multline*}
		\forall \actset_1, \actset_2, \Mt_1, \Mt_2. \\
		\configTwo{\Mt_1}{\phiassigneta{\actset_2}{\actset_1}} \To \configTwo{\Mt_2}{\Skip} \\
		\sat \Mt_1^{\actset_1} = \Mt_2^{\actset_2}
	\end{multline*}
\end{Lemma}
\begin{proof}
Recall that $\phiassigneta{\actset_2}{\actset_1}$ is just a shorthand for
assigning $v_j\in \actset_1$ to $v_i\in \actset_2$ for each $v\in \Vars$ when
$i\not=j$. Hence, for all $x\in \Vars$ such that
$\actset_1(x)\not=\actset_2(x)$, we have
$\Mt_2^{\actset_2}(x)=\Mt_1^{\actset_1}(x)$. For any other variable $x$
(such that $\actset_1(x)=\actset_2(x)$), we know its value is not updated.  Thus, we have
$ \Mt_1^{\actset_1} = \Mt_2^{\actset_2}$.
	
\end{proof}

\textbf{Proof of Theorem~\ref{thm:correctness}}: Any transformed program is
semantically equivalent to its source: 
	\begin{multline*}
	\forall \cmd, \cmdt, \Mem, \Mt, \Mem', \Mt', \actset, \actset'.\\
	\configTwo{\actset}{\cmd} \tto \configTwo{\actset'}{\cmdt} 
	\AND \configTwo{\Mem}{\cmd} \To^* \configTwo{\Mem'}{\Skip} 
	\AND \configTwo{\Mt}{\cmdt} \To^* \configTwo{\Mt'}{\Skip}
	\AND  \Mem = \Mt^{\actset} \\
	\sat \Mem' = (\Mt')^{\actset'}.
	\end{multline*}
\begin{proof} By induction on the transformation rules in the form of $ \configTwo{\actset}{\cmd} \tto \configTwo{\actset'}{\cmdt}  $.
	\begin{itemize}
		\item Case $\configTwo{\actset}{\Skip} \tto \configTwo{\actset}{\Skip} $ : trivial since no change is made to the memory $\Mem$ and the active set $\actset$.
		
		\item Case $\configTwo{\actset}{\assign{x}{e}} \tto \configTwo{\actset\Mupdate{x}{x}}{\assigneta{x}{\et}}$:
		From Lemma \ref{lem:eqexp}, we know that given $\configTwo{\Mem}{e} \evalto n$ and $\configTwo{\Mt}{\et} \evalto n'$, then $ n = n' $.  We know from the semantics rule S-Assign that $ \Mem'(x) = n $ and $ (\Mt')^{\actset'}(x) = \Mt(x) = n' $. Thus, we have $ \Mem'(x)= (\Mt')^{\actset'}(x) $. 
		For all variables other than $x$, they remain unchanged. Thus, we have $ \Mem' = (\Mt')^{\actset'} $.
		
		\item Case $\configTwo{\actset}{\bkassign{x}{e}} \tto \configTwo{\actset\Mupdate{x}{x_i}}{\assigneta{x_i}{\et}}$:
		 Similar to the previous case.
		
		\item Case $\configTwo{\actset}{\cmd_1; \cmd_2} \tto \configTwo{\actset_2}{\cmdt_1;\cmdt_2}$:
		By the induction hypothesis, given $ \configTwo{\actset}{\cmd_1} \tto \configTwo{\actset_1}{\cmdt_1}$, $ \configTwo{\Mem}{\cmd_1} \To^* \configTwo{\Mem_1}{\Skip} $ and $ \configTwo{\Mt}{\cmdt_1} \To^* \configTwo{\Mt_1}{\Skip} $, we have $ \Mem_1 = \Mt^{\actset_1}_1  $.
		By the semantics, $\configTwo{\Mem_1}{\Skip;\cmd_2} \To \configTwo{\Mem_1}{\cmd_2} \To^* \configTwo{\Mem_2}{\Skip} $ and $ \configTwo{\Mt_1}{\Skip;\cmdt_2} \To \configTwo{\Mt_1}{\cmdt_2} \To^* \configTwo{\Mt_2}{\Skip} $. By the induction hypothesis on $ \configTwo{\actset_1}{\cmd_2} \tto \configTwo{\actset_2}{\cmdt_2}$ (given $ \Mem_1 = \Mt^{\actset_1}_1$), we have $ \Mem_2 = \Mt^{\actset_2}_2  $. Thus, we have $ \Mem'= (\Mt')^{\actset'} $.
		
		\item Case $ \configTwo{\actset}{\ifcmd{e}{\cmd_1}{\cmd_2}}\! \tto\! \configTwo{\actset_3}{\ifcmd{\et}{(\cmdt_1;\phiassigneta{\actset_3}{\actset_1})}{(\cmdt_2;\actset_3=\actset_2 )} } $:
		From Lemma \ref{lem:eqexp}, we know $ e $ and $ \et $ must evaluate to some value $ n $. Thus, both evaluation must take the same branch. Without losing generality, we consider the case when $n\not = 0$.\\
		By the induction hypothesis on the transformation $ \configTwo{\actset}{\cmd_1} \tto \configTwo{\actset_1}{\cmdt_1}$, we know that given $ \configTwo{\Mem}{\cmd_1} \To^* \configTwo{\Mem_1}{\Skip} $ and 
		$ \configTwo{\Mt}{\cmdt_1} \To^* \configTwo{\Mt_1}{\Skip} $, we have $ \Mem_1 = \Mt^{\actset_1}_1  $.
	For the rest of the evaluation (	
		$ \configTwo{\Mt_1}{\Skip;\phiassigneta{\actset_3}{\actset_1}} \To \configTwo{\Mt_1}{\phiassigneta{\actset_3}{\actset_1}} \To^* \configTwo{\Mt_3}{\Skip} $), by Lemma~\ref{lem:phimem}, we have $ \Mt^{\actset_3}_3 = \Mt^{\actset_1}_1 = \Mem_1 $. Thus, we have $ \Mem' = m_1 = \Mt^{\actset_3}_3 = (\Mt')^{\actset'}$.
		
		\item $\configTwo{\actset}{\while{e}{\cmd}}\! \tto\! \configTwo{\actset_1}{\phiassigneta{\actset_1}{\actset};\while{\et}{(\cmdt;\actset_1:=\actset_2})}  $: The evaluation looks like:
			\begin{align*}
			     &\configTwo{\Mt}{\phiassigneta{\actset_1}{\actset};\while{\et}{(\cmdt;\actset_1=\actset_2)}} \\
			\To\ &\configTwo{\Mt_1}{\while{\et}{(\cmdt;\actset_1=\actset_2)}} \To ...
                        \end{align*}
By Lemma~\ref{lem:phimem}, we have $\Mt_1^{\actset_1}=\Mt^{\actset}=m$. We proceed by induction on the number of iterations being executed in the evaluation:
		\begin{itemize}

\item Base case: $0$ iteration is executed in the original program. It must be
true that $\configTwo{\Mem}{e}\evalto 0$ and $\Mem'=\Mem$. Since the transformation requires that
$\configTwo{\actset_1}{e}\tto \et$, we have
$\configTwo{\Mt_1}{\et}\evalto 0$ by Lemma~\ref{lem:eqexp}. Hence,  $0$
iteration is executed in the transformed program, and $\Mt'=\Mt_1$. Hence,
$\Mt'^{\actset_1}=\Mt_1^{\actset_1}=\Mem=\Mem'$ in this case.

\item Induction case for $ N $ iterations $ (N \geq 1) $: The evaluation of original $ \While $ looks like:
			\begin{align*} 
                             &\configTwo{\Mem}{\while{e}{\cmd}} \\
			\To\ &\configTwo{\Mem}{\ifcmd{e}{(c;\while{e}{\cmd})}{\Skip}} \\
			\To\ &\configTwo{\Mem}{c;\while{e}{\cmd}} \\
			\To^* &\configTwo{\Mem_1}{\Skip;\while{e}{\cmd}} \\
			\To\ &\configTwo{\Mem_1}{\while{e}{\cmd}} \\
			\To\ &\dots N-1 \text{ iterations} \dots 
                        \end{align*}
			The evaluation of the transformed $\While$ looks like:
			\begin{align*}
			     &\configTwo{\Mt_1}{\while{\et}{(\cmdt;\actset_1:=\actset_2)}} \\
			\To\ &\configTwo{\Mt_1}{\ifcmd{\et}{(\cmdt;\actset_1:=\actset_2;\while{\et}{(\cmdt;\actset_1:=\actset_2)})}{\Skip}} \\
			\To\ &\configTwo{\Mt_1}{\cmdt;\actset_1:=\actset_2;\while{\et}{(\cmdt;\actset_1:=\actset_2)}} \\
			\To^* &\configTwo{\Mt_2}{\Skip;\actset_1:=\actset_2;\while{\et}{(\cmdt;\actset_1:=\actset_2)}} \\
			\To\ &\configTwo{\Mt_2}{\actset_1:=\actset_2;\while{\et}{(\cmdt;\actset_1:=\actset_2)}}\\
			\To\ &\configTwo{\Mt_3}{\while{\et}{(\cmdt;\actset_1:=\actset_2)}} \To ...
                        \end{align*}

Here, we know that the transformed program must take the ``if'' branch due to
the same argument as in the base case.  Since $\Mt_1^{\actset_1}=m$, by
(structural) induction hypothesis on $c$ and the transformation rule which
requires $\configTwo{\actset_1}{\cmd} \tto \configTwo{\actset_2}{\cmdt}$, we
have $\Mem_1=\Mt_2^{\actset_2}$. By Lemma~\ref{lem:phimem}, we have
$\Mt_3^{\actset_1}=\Mt_2^{\actset_2}=\Mem_1$. By the induction hypothesis when
the original program runs for $N-1$ iterations, we have
$\Mt'^{\actset_1}=\Mem'$ as desired.

		\end{itemize}

	\end{itemize}
\end{proof}

\clearpage
\section{Soundness}
\label{appendix:soundness}

We need more definitions before showing the soundness of our analysis. We first
use a distinguished label $L$ (``low'') to define what is observable to the low
observer. Since the lemmas and theorems are valid regardless of what level $L$
is, the propositions proved hold for any label $\lbl$ in the security lattice.  




As discussed in Section~\ref{sec:soundness}, we prove the soundness based on
the erasure semantics in Figure~\ref{fig:erasure}. The connection between the
erasure and the standard semantics is established by the following lemma:
\begin{Lemma}
\label{lem:erasurecorrectness}
\begin{multline*}
∀\cmdt,\Mt,\Mt_1,\Mt_2, \actset.~\configTwo{\cmdt}{\Mt}\To^* \configTwo{\Skip}{\Mt_1} ∧
    \configTwo{\cmdt}{\Mt}\ERTo{\actset}^* \configTwo{\Skip}{\Mt_2} \\
  ⇒ ∀v\in \actset.~\Mt_1(v)=\Mt_2(v)
\end{multline*}
\end{Lemma}
\begin{proof}
We note that the erasure semantics only changes the values of dead variables.
Hence, the result is trivial given that the live variable analysis is correct.
\end{proof}

Next, we prove that any well-typed target program under the erasure semantics
satisfies the noninterference property. To simplify notation, we will use $\To$
instead of $\ERTo{\actset}$ hereafter.  To prove soundness, we extend the
language syntax and semantics with explicitly marked high values and commands.
Memory is extended to track high values as well. The extension is useful since
the low equivalence relation we defined earlier corresponds to the equivalence
relation on the marked memories. 

By showing the completeness of the extended language, all interesting proof are
then conducted on the extended language. To do that, we first prove several
useful lemmas, and then show the type system enforces noninterference.

\subsection{Extended Language}

\paragraph{Extended syntax}

The extended syntax is shown in Figure~\ref{fig:ext:syntax}.
We augment memories to map high variables to bracketed results. In a
similar way, syntax is augmented to include bracketed results, and
bracketed commands. Intuitively, bracketed results represent values
from high memory,
and bracketed commands represent commands executed in a high pc
context (such as in a branch with a high guard). 

\begin{figure}
\begin{minipage}{0.3\textwidth}
\centering
\begin{align*} 
e ::=\; & \dots\ |\ [n] \\
c ::= \; & \dots\ |\ [c]
\end{align*} 
\caption{Extended Syntax}
\label{fig:ext:syntax} 
\end{minipage}
\begin{minipage}{0.65\textwidth}
\centering
\begin{mathpar} 
\inferrule{}{n \eq n} \quad \inferrule{}{[n_1] \eq [n_2]} 

\quad  
\Mem_1 \eq \Mem_2 \implies \forall x. \Mem_1(x) \eq \Mem_2(x)
 
\inferrule{}{\cmd \eq \cmd} \quad
\inferrule{\cmd_1 \eq \cmd_3 \and \cmd_2\eq
\cmd_4}{\cmd_1;\cmd_2 \eq
\cmd_3;\cmd_4} 
\quad \inferrule{}{[\cmd_1]\eq
[\cmd_2]}
\end{mathpar}
\caption{Equivalence on Memories and Commands} 
\label{fig:equivalence} 
\end{minipage}

\end{figure} 

\begin{figure}
\centering
\begin{mathpar} 
	\inferrule{ }{\configTwo{\Mem}{[n]} \evalto {[n]}} 
	\and 
	\inferrule{\configTwo{\Mem}{e_1} \evalto {[n_1]} \and 
		\configTwo{\Mem}{\expr_2} \evalto {n_2} \and n = n_1\ \op\ n_2 } 
	{ \configTwo{\Mem}{e_1\ \op\ e_2} \evalto {[n]} } 
	\and
	\inferrule{\configTwo{\Mem}{e_1} \evalto {n_1} \and 
		\configTwo{\Mem}{e_2} \evalto {[n_2]} \and n = n_1\ \op\ n_2 } 
	{ \configTwo{\Mem}{e_1\ \op\ e_2} \evalto {[n]} } 
	\and
	\inferrule{\configTwo{\Mem}{e_1}\evalto {[n_1]} \and 
		\configTwo{\Mem}{e_2} \evalto {[n_2]} \and n = n_1\ \op\ n_2 } 
	{ \configTwo{\Mem}{e_1\ \op\ e_2} \evalto {[n]} }
\end{mathpar} 
\caption{Extended Semantics: Expressions}
\label{fig:ext:expsem}

\begin{mathpar} 
		\switch(\Mem, x, \eta)(\varx) =
		\begin{cases}
			0, &  x \in \free(\varx) ∧ x'\not\in \liveset{\actset}(\eta) ∧ \labelof{\varx}{\Mem} \absleq L \\
			[0], & x \in \free(\varx) ∧ x'\not\in \liveset{\actset}(\eta) ∧ \labelof{\varx}{\Mem} \not \absleq L  \\
			\Mem(\varx), & \text{otherwise}\\
		\end{cases}\\
	\inferrule[S-Skip1]
	{ }{\configTwo{\Mem}{[\Skip]} \To \configTwo{\Mem}{\Skip}} 
	\and 
	\inferrule[S-Bracket]
	{\configTwo{\Mem}{\cmd} \To\configTwo{\Mem'}{\cmd'} }
	{\configTwo{\Mem}{[\cmd]} \To\configTwo{\Mem'}{[\cmd']} } 
	\and
	\inferrule[S-Asgn1] 
	{\configTwo{\Mem}{e} \evalto {n} \\ \labelof{x}{\Mem}\sqsubseteq L \\ \Mem' = \Mem\Mupdate{x}{n}}
	{\configTwo{\Mem}{\assigneta{x}{e}} \To \configTwo{\switch(\Mem', x, \eta)}{\Skip} }
	\and 
	\inferrule[S-Asgn2] 
	{\configTwo{\Mem}{e} \evalto {n} \\ \labelof{x}{\Mem}\not\sqsubseteq L \\ \Mem' = \Mem\Mupdate{x}{[n]} }
	{\configTwo{\Mem}{\assigneta{x}{e}} \To \configTwo{\switch(\Mem', x, \eta)}{\Skip} } 
	\and
	\inferrule[S-Asgn3] 
	{\configTwo{\Mem}{e} \evalto {[n]} \\ \Mem' = \Mem\Mupdate{x}{[n]}}
	{\configTwo{\Mem}{\assigneta{x}{e}} \To \configTwo{\switch(\Mem', x, \eta)}{\Skip} } 
	\and 
	\inferrule[S-If3]
	{\configTwo{\Mem}{e} \evalto {[n]}  \\ n \neq 0}
	{\configTwo{\Mem}{\ifcmd{e}{c_1}{c_2}} \To \configTwo{\Mem}{[c_1]}} 
	\and
	\inferrule[S-If4]
	{\configTwo{\Mem}{e} \evalto {[n]} \\ n = 0}
	{\configTwo{\Mem}{\ifcmd{e}{c_1}{c_2}} \To \configTwo{\Mem}{[c_2]}} 
\end{mathpar}
\caption{Extended Semantics: Commands} 
\label{fig:ext:cmdsem}
\end{figure}

\paragraph{Extended Semantics}

The operational semantics is augmented to propagate brackets, as shown in
Figure~\ref{fig:ext:expsem},~\ref{fig:ext:cmdsem}.
All rules are extensions to the original grammar except that
\ruleref{S-Asgn} is
split into three rules: \ruleref{S-Asgn1}, \ruleref{S-Asgn2},
\ruleref{S-Asgn3}. Moreover, the erasure semantics for the extended language
also adds brackets when needed.  All rules with brackets work the same way as
the normal rules from computational perspective. Brackets are just syntactic
markers.

\begin{figure}
\begin{minipage}[b]{0.5\textwidth}
	\centering
\begin{mathpar}
%
\inferrule*[right=T-BracketExp]
  {\ell \not \sqsubseteq L}
  {\G \proves [n] : \ell} 
\end{mathpar}
\caption{Extended Typing Rules: Expressions}
\label{fig:typing:exp}
\end{minipage}
\begin{minipage}[b]{0.5\textwidth}
	\centering
\begin{mathpar} 
\inferrule[T-BracketCmd]
  {\G, \tau \proves \cmd \and \pc \sqsubseteq \tau \and  \tau \not \sqsubseteq L}
  {\G, \pc \proves [\cmd]}
\end{mathpar}
\caption{Extended Typing Rules: Commands} 
\label{fig:ext:typing} 
\end{minipage}
\end{figure}

The extended language requires extra typing rules, shown in
Figure~\ref{fig:typing:exp}.  Rule \ruleref{T-BracketExp} treats bracketed
expression as high. Note that this rule requires a security level $\ell$, which
by definition cannot depend on any program state.  Rule~\ruleref{T-BracketCmd}
in Figure~\ref{fig:ext:typing} is given to support the soundness proof.
Bracketed command should be type-checked under a pc label that is not bounded by
$L$ in the type system. 

\paragraph{Equivalence on Memories and Commands}

We define the equivalence of memories and commands up to label $L$
as in Fig.~\ref{fig:equivalence}. 
Intuitively, bracketed memory and commands are indistinguishable.  An
equivalence relation $\eq$ is defined on memories such that $\Mem_1 \eq \Mem_2$ 
if
and only if they agree on all low variables, and the rest of variables are all
high (with brackets).

The type system enforces an important invariant on the memory: a variable holds
bracketed value if and only if the security level of that variable is high.
This is formalized as follows:

\begin{Definition}[Well-Formedness] 
A variable $ x $ is well-formed under memory $\Mem$, denoted as $
\wellform_{\Mem} x $ if the following condition holds:
	\[	\labelof{x}{m} \not \LEQ L \iff \exists n. \configTwo{\Mem}{x}\evalto [n] \]

A memory $\Mem$ is well-formed, denoted as $ \wellform \Mem $ if all variables
are well formed under $\Mem$:
	\[ \wellform \Mem \iff ( \forall x \in \Mem . \wellform_{\Mem} x) \]
\end{Definition}

\paragraph{Completeness of the Extended Language}

It is then clear that for any two low-equivalent standard memories $\Mem_1$,
$\Mem_2$, there are augmented memories, simply by putting brackets for high
variables, that agree with standard memories on all low variables; and vise
versa. Hence, the completeness result (Lemma~\ref{lem:completeness}) justifies
the noninterference result in the unextended language, by showing that starting
from any $\Mem_1 \eq \Mem_2$ that are both well-formed, the resulting memories
are still equivalent in the augmented language.  

Completeness means that every step in the new semantics can be performed in the
unextended semantics (maybe with removal of brackets) and vice versa.
More formally, given that $c$ is a command in the extended
language, let us use the notation of $\lfloor c \rfloor$ to
denote removal of all brackets from $c$ in the
obvious way, yielding a command from the original language. Similarly, we
define $\lfloor m \rfloor$ to convert memory. Completeness can be
expressed as the following lemma. 

\begin{Lemma}[Completeness of the Extended Language] 
	\label{lem:completeness}
	\[
	\G, \pc \proves c \AND \configTwo{\lfloor m \rfloor}{\lfloor c
		\rfloor} \To^*
	\configTwo{\Mem'}{\Skip} \implies \exists m''.\
	\configTwo{\Mem}{c}\To^*
	\configTwo{\Mem''}{\Skip}~\AND~ m'=\lfloor m'' \rfloor
	\]
\end{Lemma}

\begin{proof} By rule induction on each evaluation step.  \end{proof}

\subsection{Soundness Proof}
Next, we prove the soundness of the type system on the extended language
language with the erasure semantics. We first introduce a couple of useful
lemmas.

\begin{Lemma} 
	\label{lem:conserve} 
	The type comparison is conservative: 
	\[∀m, \Type_1, \Type_2~.~\Type_1 \sqsubseteq \Type_2 \implies
	\valueof{\Type_1}{m} \sqsubseteq \valueof{\Type_2}{m}\] 
\end{Lemma}
\begin{proof} 
Clear from the lifted definition of $\sqsubseteq$ on (dependent) labels.
\end{proof}


\begin{Lemma} 
	\label{lowexp} 
	Low expressions always evaluate to
	ordinary integers (without brackets) under well-formed memory: 
	\[ \wellform \Mem \AND  \G \proves e
	: \tau \AND \valueof{\tau}{\Mem} \sqsubseteq L \implies \exists n. \configTwo{\Mem}{e} \evalto n\] 
\end{Lemma}

\begin{proof} By induction on the structure of the expression $ e  $:
	\begin{itemize}
		\item Case $ e = n $: trivial.
		\item Case $ e = [n] $: contradiction to the typing rule \ruleref{T-BracketExp}.
		\item Case $ e = x $: $ x $  could be either low or high.
		\begin{itemize}
			\item  Case $ x $ is low ($\labelof{x}{m} \sqsubseteq L $ ): clear from
			the definition of $ \wellform \Mem $. 
			\item  Case $ x $ is high ($\labelof{x}{m} \not\sqsubseteq L $ ):
			contradiction to the assumption $  \valueof{\tau}{m} \sqsubseteq L $.
		\end{itemize}
		\item Case $ e = e_1 ~\op ~ e_2 $:
		From $ \G \proves e : \tau $, we can infer that $ \G \proves e_1:\tau_1$, $Γ\proves e_2:\tau_2 $ and $ \tau = \tau_1 \join \tau_2$. By Lemma~\ref{lem:conserve}, $\valueof{\tau_1}{\Mem} \sqsubseteq L$ and $\valueof{\tau_2}{\Mem}\sqsubseteq L$. Hence
		by induction hypothesis, $ \exists n_1, n_2. \configTwo{\Mem}{e_1} \evalto n_1, \configTwo{\Mem}{e_2} \evalto n_2 $. 
		Thus, $ \configTwo{\Mem}{e} \evalto n $, where $ n = n_1 ~\op~ n_2 $ by the semantics.
	\end{itemize}
	
\end{proof}

\begin{Lemma}[PC Subsumption] 
	\label{lem:pcsub} 
	\[\G, \pc \proves \cmd \AND \pc' \sqsubseteq  \pc \implies \G, \pc' \proves \cmd\] 
\end{Lemma} 
\begin{proof}  By rule induction on the typing derivation for $ \cmd $ : 
	\begin{itemize}
		\item Case $ \Skip $ : From typing rule \ruleref{T-Skip}, we know that $ \G , \pc' \proves \Skip $ for any $ \pc' $.

		\item Case $ \assigneta{x}{e} $ : From typing rule \ruleref{T-Assign}, we know that $ \G  \proves e : \tau $ and $  \valid{\hypo(\before{\eta}) \sat \Type \join \pc \absleq \G(x)} $ and $ \forall v \in \liveset{\actset}{(\after\eta)}.  x \not \in \free(\G(v)) $. Since $ \pc' \sqsubseteq \pc $, we have $ \tau \join \pc' \sqsubseteq \tau \join \pc \sqsubseteq \G(x) $. Thus, we have $ \valid{\hypo(\before \eta) \sat \tau \join \pc' \sqsubseteq \G (x)} $ and there is no change to the other conditions. So we can derive $  \G, \pc' \proves \assigneta{x}{e} $.

		\item Case $ \cmd_1;\cmd_2 $ : 
		From typing rule \ruleref{T-Seq}, we know that $ \G, \pc \proves \cmd_1$ and $ \G, \pc \proves \cmd_2  $. 
		By the induction hypothesis, we have $ \G, \pc' \proves \cmd_1$ and $ \G, \pc' \proves \cmd_2  $. Thus, we can derive $ \G, \pc' \proves \cmd_1;\cmd_2 $.

		\item Case $ \ifcmd{e}{\cmd_1}{\cmd_2} $ : 
		From typing rule \ruleref{T-If}, we know that $ \G \proves e:\tau $, $~ \G, \pc\join\tau \proves \cmd_1   $ and $  \G, \pc\join\tau \proves \cmd_2  $. Since $ \pc' \sqsubseteq \pc $, we have $ \tau \join \pc' \sqsubseteq \tau \join \pc $. So by the induction hypothesis, we have $  \G, \pc'\join\tau \proves \cmd_1$ and $  \G, \pc'\join\tau \proves \cmd_2  $. Hence, $ \G, \pc' \proves \ifcmd{e}{\cmd_1}{\cmd_2}$.


		\item Case $ \while{e}{\cmd}$ : 
		From typing rule \ruleref{T-While}, we know that $ \G \proves e :\tau $, $~ \G, \pc\join\tau \proves \cmd $. Since $ \pc' \sqsubseteq \pc $, we have $ \tau \join \pc' \sqsubseteq \tau \join \pc $. So by the induction hypothesis, we have $ \G, \pc'\join\tau \proves \cmd$.  Hence, $ \G, \pc' \proves\while{e}{\cmd}$.
				
	\end{itemize}
\end{proof}


\begin{Lemma}[Preservation] 
	\label{lem:preservation}
	\begin{multline*}
	\G, \pc \proves \cmd 
	\AND ~ \configTwo{\Mem_1}{\cmd} \To\configTwo{\Mem_2}{\cmd'} 
	\AND~ \wellform \Mem_1 \\ 
	\implies \wellform \Mem_2 \AND~ \G, \pc \proves \cmd'
	\end{multline*}
\end{Lemma}

\begin{proof} 
	By rule induction on the evaluation rules $\configTwo{\Mem_1}{\cmd} \To\configTwo{\Mem_2}{\cmd'}$: 
	\begin{itemize}
		\item Case $ \configTwo{\Mem_1}{[\Skip]} \To\configTwo{\Mem_1}{\Skip} $: trivial.
		
		\item Case $ \configTwo{\Mem_1}{\Skip;\cmd} \To\configTwo{\Mem_1}{\cmd} $ : Trivial since $\Mem$ does not change  and $ \G , \pc \proves\cmd$ is required in rule~\ruleref{T-Seq}. 
		
		\item Case $ \configTwo{\Mem_1}{\cmd_1; \cmd_2} \To \configTwo{\Mem_2}{\cmd_1';\cmd_2} $: 
		From the assumption, we have $ \configTwo{\Mem_1}{\cmd_1} \To \configTwo{\Mem_2}{\cmd_1'}$. 
		From typing rule \ruleref{T-Seq}, we have $ \G , \pc \proves \cmd_i, i\in\{1,2\}$. 
		So by the induction hypothesis, we have $ \G , \pc \proves \cmd_1'$ and $ \wellform  \Mem_2 $. 
		Hence we can derive $ \G , \pc \proves \cmd_1';\cmd_2 $.
		
		\item Case $\configTwo{\Mem_1}{[\cmd]} \To\configTwo{\Mem_2}{[\cmd']}$: 
		From typing rule \ruleref{T-BracketCmd}, we know that there exists some label $\tau$ such that $\tau \not \sqsubseteq L ∧ \pc \sqsubseteq \tau $ and $ \G , \tau \proves \cmd $.
		From the assumption, we know that $\configTwo{\Mem_1}{\cmd} \To\configTwo{\Mem_2}{\cmd'}$.
		By induction hypothesis, we have $ \G , \tau \proves \cmd'$ and $ \wellform  m_2 $. 
		Thus, we can derive $ \G , \pc \proves [\cmd'] $ .
		
		\item Case $\configTwo{\Mem_1}{\assigneta{x}{e}} \To \configTwo{\switch(\Mem', x, \eta)}{\Skip}$:
		We have $ \G , \pc \proves \Skip $ trivially. Next, we prove that $ \wellform \Mem_2 $ (in this case, $\Mem_2$ is $\switch(\Mem', x, \eta)$) by showing that every variable $\varx$ is well-formed under $\Mem_2$. To do so, we first note that $ \labelof{\varx}{\Mem_2} = \labelof{\varx}{\Mem'}$ since there is no chain of dependence or self dependency. 

		\begin{itemize}
			\item Case $  x \not \in \free(\G(\varx)) $: We first show that the level of $x'$ does not change after the assignment. Since $ \varx $ does not depend on $ x $, we have $ \labelof{\varx}{\Mem_1} = \labelof{\varx}{\Mem'} $. Further, we can infer that $ \labelof{\varx}{\Mem_2} = \labelof{\varx}{\Mem'} = \labelof{\varx}{\Mem_1} $. 

By the definition of $ \switch $, we have $ \Mem_2(\varx) = \Mem'(\varx) $. So
if $ \varx \not = x $, we have $ \wellform_{\Mem_2} \varx $ trivially since
neither its value (since $\Mem'(\varx)=\Mem(\varx)$) nor its type is changed
after the assignment; otherwise, if $ \varx = x $ we have three cases:

			\begin{itemize}
				\item Case S-Asgn1: We have $ \Mem_2(x) = \Mem'(x) = \Mem_1\Mupdate{x}{n}(x)=n $ and $ \labelof{x}{\Mem_2} = \labelof{x}{\Mem_1} \sqsubseteq L $. Thus, $ \wellform_{\Mem_2} x $.
				\item Case S-Asgn2: We have $ \Mem_2(x) = \Mem'(x) = \Mem_1\Mupdate{x}{[n]}(x)=[n] $ and $ \labelof{x}{\Mem_2} = \labelof{x}{\Mem_1} \not \sqsubseteq L $. Thus, we can derive $ \wellform_{\Mem_2} x $ since $ x $ is high and is given a bracketed value.
				\item Case S-Asgn3: We have $ \Mem_2(x) = \Mem'(x) = \Mem_1\Mupdate{x}{[n]}(x)=[n] $. We also know from the typing rule \ruleref{T-Assgn} that $ \G \proves e : \tau, ~ \valid { \hypo (\before \eta) \sat \tau \join \pc \sqsubseteq \G(x)} $. By assumption, $\configTwo{\Mem_1}{e} \evalto {[n]}$. By Lemma~\ref{lowexp}, $\valueof{\tau}{\Mem_1}\not\LEQ L$. Hence, $\labelof{x}{\Mem_1}\not\LEQ L$ due to Lemma~\ref{lem:conserve}. Thus, we have $ \labelof{x}{\Mem_2} = \labelof{x}{\Mem_1} \not \sqsubseteq L $. So $ \wellform_{\Mem_2} x $  since $ x $ is high and is given a bracketed value.
			\end{itemize}
			Thus, in all three cases, we have $ \wellform_{\Mem_2} \varx $.
			\item Case $  x \in \free(\varx) $: by the definition of $ \switch $, $ \varx $ is erased to $ 0 $ or $ [0] $ according to $ \labelof{x'}{\Mem'} $. We already showed $ \labelof{\varx}{\Mem'} = \labelof{\varx}{\Mem_2} $. Thus, we have $ \wellform_{\Mem_2} \varx $.
		\end{itemize}
		
		\item Case $ \configTwo{\Mem_1}{\ifcmd{e}{\cmd_1}{\cmd_2}} \To\configTwo{\Mem_1}{\cmd_1} $ and $ \configTwo{\Mem_1}{\ifcmd{e}{\cmd_1}{\cmd_2}} \To \configTwo{\Mem_1}{\cmd_2} $ : 
		$\proves \Mem_2$ is trivial since $\Mem_2=\Mem_1$. For types, from the typing rule \ruleref{T-If}, we have $ \G \proves e : \tau$,  $ \G , \pc \join \tau \proves \cmd_1 $ and $\G , \pc \join \tau \proves \cmd_2 $.
		Since $ \pc \sqsubseteq \pc \join \tau $, we can derive from Lemma \ref{lem:pcsub} that  $ \G , \pc \proves \cmd_1 $ and $ \G , \pc \proves \cmd_2 $.
		
		\item Case $ \configTwo{\Mem_1}{\ifcmd{e}{\cmd_1}{\cmd_2}} \To\configTwo{\Mem_1}{[\cmd_1]} $, and $ \configTwo{\Mem_1}{\ifcmd{e}{\cmd_1}{\cmd_2}} \To\configTwo{\Mem_1}{[\cmd_2]} $ : $\proves \Mem_2$ is trivial since $\Mem_2=\Mem_1$. For types, from the typing rule \ruleref{T-If}, we have $ \G \proves e : \tau$,  $ \G , \pc \join \tau \proves \cmd_1 $ and $ \G , \pc \join \tau \proves \cmd_2 $. 
		From the assumption, we have $ \configTwo{\Mem_1}{e} \evalto [n] $. By Lemma~\ref{lowexp}, $\valueof{\tau}{\Mem_1}\not\LEQ L$. Hence, $\labelof{\pc\join\tau}{\Mem_1}\not\LEQ L$ due to Lemma~\ref{lem:conserve}. Thus, we can derive $ \G , \pc  \proves [\cmd_1] $ and $ \G , \pc \proves [\cmd_2] $ by using $\pc\join\tau$ as the ``$\tau$'' in rule \ruleref{T-BracketCmd}.  
		
		
		\item Case $ \configTwo{\Mem_1}{\while{e}{\cmd}} \To\configTwo{\Mem_1}{\ifcmd{e}{(\while{e}{\cmd})}{\Skip}} $ : \\
		$\proves \Mem_2$ is trivial since $\Mem_2=\Mem_1$. By the typing rule \ruleref{T-While}, we have two assumptions $ A = \G \proves e : \tau$ and $B = \G, \Type \join \pc \proves \cmd $. So the program after evaluation can be typed as follows:
		\begin{mathpar}
				\inferrule	{	A 
					\quad 
					\inferrule	{A	
						\quad    B	
					}{	 \G, \Type \join \pc \proves \while{e}{\cmd} }
					\quad 
					\inferrule	{ }
					{ \G, \Type \join \pc \proves \Skip 	}
				}{	\G, \pc \proves \ifcmd{e}{(\while{e}{\cmd})}{\Skip} }
			\end{mathpar}
	\end{itemize}  
\end{proof}

\begin{Lemma}[High-Step] 
	A command that type-checks in a high-pc context only modifies high variables. 
	\label{lem:high-step} 
	\[\forall \Mem_1, \Mem_2, \cmd, \pc~.~
	\valueof{\pc}{\Mem_1} \not\sqsubseteq L ~
	\AND~ \G, \pc \proves \cmd~
	\AND~ \wellform \Mem_1~
	\AND~ \configTwo{\Mem_1}{\cmd} \To \configTwo{\Mem_2}{\cmd'} 
	\implies \Mem_1 \eq \Mem_2\]
\end{Lemma}

\begin{proof}
	By induction on evaluation rules $ \configTwo{\Mem_1}{\cmd} \To \configTwo{\Mem_2}{\cmd'} $:
	\begin{itemize}
		\item Cases $ \configTwo{\Mem_1}{[\Skip]} \To\configTwo{\Mem_1}{\Skip}$,$ \configTwo{\Mem_1}{\Skip;\cmd} \To\configTwo{\Mem_1}{\cmd} $,\\
		$ \configTwo{\Mem_1}{\ifcmd{e}{\cmd_1}{\cmd_2}} \To\configTwo{\Mem_1}{\cmd_1} $,  
		$ \configTwo{\Mem_1}{\ifcmd{e}{\cmd_1}{\cmd_2}} \To\configTwo{\Mem_1}{\cmd_2} $,\\ 
		$ \configTwo{\Mem_1}{\ifcmd{e}{\cmd_1}{\cmd_2}} \To\configTwo{\Mem_1}{[\cmd_1]}$, 
		$ \configTwo{\Mem_1}{\ifcmd{e}{\cmd_1}{\cmd_2}} \To\configTwo{\Mem_1}{[\cmd_2]} $, \\
		$ \configTwo{\Mem_1}{\while{e}{\cmd}} \To\configTwo{\Mem_1}{\ifcmd{e}{(\while{e}{\cmd})}{\Skip}} $: \\
		Trivial since $ \Mem_2=\Mem_1$.
	
		\item Case $ \configTwo{\Mem_1}{\cmd_1; \cmd_2} \To \configTwo{\Mem_2}{\cmd_1';\cmd_2} $ : 
		From the type rule \ruleref{T-Seq}, we have $ \G, \pc \proves \cmd_1 $. By induction hypothesis on the assumption $\configTwo{\Mem_1}{\cmd_1} \To\configTwo{\Mem_2}{\cmd_1'}$ , we have $ \Mem_1 \eq \Mem_2 $. 
		
		\item Case $\configTwo{\Mem_1}{[\cmd]} \To\configTwo{\Mem_2}{[\cmd']}$ :
		From the evaluation rule, we have  $\configTwo{\Mem_1}{\cmd} \To\configTwo{\Mem_2}{\cmd'}$.
		From the typing rule, we have $ \G , \tau \proves\cmd$ for some $\tau$ such that $\pc\LEQ \tau$. Since by assumption, $\valueof{\pc}{\Mem_1}\not\LEQ L$, we have $\valueof{\tau}{\Mem_1}\not\LEQ L$ by Lemma~\ref{lem:conserve}. Hence,
		by induction hypothesis on the evaluation assumption, we have $ \Mem_1 \eq \Mem_2 $.
		
		\item Case$\configTwo{\Mem_1}{\assigneta{x}{e}} \To \configTwo{\switch(\Mem', x)}{\Skip}$ : 
		First, we can infer from the type rule that $\G\proves e:\tau$, $\valid { \hypo (\before \eta) \sat \tau \join \pc \sqsubseteq \G(x)} $. Due to the correctness of predicate generator, we have $\tau \join \pc \sqsubseteq \G(x)$. Since $ \valueof{\pc}{\Mem_1} \not \absleq L $, we have $  \labelof{x}{\Mem_1} \not \sqsubseteq L $ by Lemma~\ref{lem:conserve}. From assumption $ \wellform \Mem_1 $, we know $ x $ must hold a bracketed value in $ \Mem_1 $. We also know that S-Asgn1 can not be applied since it requires $ \labelof{x}{\Mem_1} \sqsubseteq L $. When S-Asgn2 or S-Asgn3 is applied, we have $ \Mem' = \Mem_1\Mupdate{x}{[n]} $. Since there is no self-dependence, $\Mem_2(x)=\Mem'(x)=[n]$. So $\Mem_2(x)\eq \Mem_1(x)$. Next, we show that for variable $ \varx\not=x $, we have $ \Mem_1(\varx) \eq \Mem_2(\varx) $:
		\begin{itemize}
			\item Case $  x \not \in \free(\varx) $: by definition of $ \switch $, $\Mem_2(\varx)=\Mem'(\varx)=\Mem_1(\varx)$. Hence, $ \Mem_1(\varx) \eq \Mem_2(\varx)$. 
			\item Case $  x \in \free(\varx) $: by definition of $ \switch $, $ \varx $ is erased to $ 0 $ or $ [0] $ according to $ \labelof{\varx}{\Mem'} $. We already know that $ \labelof{x}{\Mem_1} \not \sqsubseteq L $. Since there is no self-dependence, $ \labelof{x}{\Mem'} \not \sqsubseteq L $. Recall that $  x \in \free(\varx) $ implies $\G(x)\LEQ \G(x')$. Hence, by Lemma~\ref{lem:conserve}, $ \labelof{\varx}{\Mem_1} \not \sqsubseteq L $ and $ \labelof{\varx}{\Mem'} \not \sqsubseteq L $. By assumption $\proves \Mem_1$, $\varx$ holds a bracketed value in $\Mem_1$. Moreover, by the erasure semantics, $x'$ is erased to [0] after the assignment. So $\Mem_1(\varx)\eq \Mem_2(\varx)$.
		\end{itemize}		
		

	\end{itemize}
	
\end{proof}


\begin{Lemma} 
	\label{lem:expeq} 
	\[\forall \Mem_1, \Mem_2. ~
	\Mem_1 \eq \Mem_2  ∧
	\configTwo{\Mem_1}{e} \evalto v_1 \\ 
	\implies \exists v_2. \configTwo{\Mem_2}{e} \evalto v_2 \AND v_1 \eq v_2\] 
\end{Lemma}

\begin{proof} By rule induction on the structure of expression $ e $. 
	\begin{itemize}
		\item Case $ e = n $: trivial since we have $ \configTwo{\Mem}{e}\evalto n $ for any memory, so $ v_1 = v_2 = n $.
		\item Case $ e = [n] $: trivial since we have $ \configTwo{\Mem}{e}\evalto [n] $ for any memory, so $ v_1 = v_2 = [n] $.
		\item Case $ e = x $: 
		From $ \Mem_1 \eq \Mem_2 $, we have $ \Mem_1(x) = \Mem_2(x)$, thus, $ v_1 \eq v_2 $.
		\item Case $ e = e_1 ~\op ~ e_2  $: 
		By the induction hypothesis, we have $ \configTwo{\Mem_1}{e_1} \evalto v_3, \configTwo{\Mem_2}{e_1} \evalto v_3', \configTwo{\Mem_1}{e_2} \evalto v_4, \configTwo{\Mem_2}{e_2} \evalto v_4', $ and $ v_3 \eq v_3', v_4 \eq v_4'$. 
		\begin{itemize}
			\item If both $ v_3, v_4 $ are non-bracketed values, then $v_3=v_3'$ and $v_4=v_4'$. Result is trivial.
			\item If at least one of $ v_3 $ and $ v_4 $ holds bracketed value, we know that at least one of $v_3'$ and $ v_4' $ holds a bracketed value. From the semantics we know $ e $ must be evaluated to bracketed values under both $\Mem_1$ and $\Mem_2$. Thus, $ v_1 \eq v_2 $.
		\end{itemize}

	\end{itemize}
\end{proof}


\begin{Lemma}[Unwinding] 
	\label{lem:unwinding} 
	\begin{multline*}
∀\cmd_1, \cmd_2, \Mem_1, \Mem_2, \Mem_3, \Mem_4.~\\
	\proves \cmd_1 
	∧ \proves \cmd_2  
	∧ \cmd_1 \eq \cmd_2 
	∧ \wellform \Mem_1 
	∧ \wellform \Mem_2 
	∧ \Mem_1 \eq \Mem_2 \\
	∧ \configTwo{\Mem_1}{\cmd_1} \To \configTwo{\Mem_3}{\cmd_3} \\
	\implies (\exists \cmd_4, \Mem_4.~\cmd_4 \eq \cmd_3
				∧ \configTwo{\Mem_2}{\cmd_2} \To^* \configTwo{\Mem_4}{\cmd_4} 
				∧ \Mem_3 \eq \Mem_4) 
			\lor (\configTwo{\Mem_2}{\cmd_2} \Uparrow \AND \exists \cmd.~\cmd_2 =[\cmd])
	\end{multline*} 
\end{Lemma}

\begin{proof} By rule induction on $\configTwo{\Mem_1}{\cmd_1} \To\configTwo{\Mem_3}{\cmd_3}$.  
	
	\begin{itemize}
		
		\item Case $ \configTwo{\Mem_1}{[\Skip]} \To \configTwo{\Mem_1}{\Skip} $:
		From $ \cmd_1 \eq \cmd_2 $, we know $ \cmd_2 $ has the form of $ [\cmd_5]$, for some $ \cmd_5 $. If $ \cmd_2 $ diverges, we are done with $\cmd= \cmd_5$.
		Otherwise, we have $ \configTwo{\Mem_2}{[\cmd_5]} \To^* \configTwo{\Mem_4}{[\Skip]} \To \configTwo{\Mem_4}{\Skip}$. By induction on the number of steps using Lemma \ref{lem:high-step}, we have $ \Mem_4 \eq \Mem_2 \eq \Mem_1 $. So we can choose $ \cmd_4 = \Skip $.
		
		\item Case $ \configTwo{\Mem_1}{[\cmd_5]} \To\configTwo{\Mem_3}{[\cmd_5']}  $:
		From $ \cmd_1 \eq \cmd_2 $, we know $ \cmd_2 $ has the form of $ [\cmd_6]$ for some $\cmd_6$. Hence, $ \cmd_2=[\cmd_6] \eq [\cmd_5']  $. Moreover, by Lemma~\ref{lem:high-step}, $\Mem_3\eq \Mem_1$. Hence, $\Mem_2 \eq \Mem_3 $.
		Therefore, we can choose $\cmd_4=\cmd_2$ and make zero step under $\Mem_2$.
		
		\item Case $ \configTwo{\Mem_1}{\Skip;\cmd_5} \To\configTwo{\Mem_1}{\cmd_5} $:
		Command $ \cmd_2 $ must also have the form of $ \Skip;\cmd_6 $ where $ \cmd_5 \eq \cmd_6 $. So $ \configTwo{\Mem_2}{\Skip;\cmd_6} \To\configTwo{\Mem_2}{\cmd_6} $ preserves the equivalence on memory and command as required.
		
		\item Case $ \configTwo{\Mem_1}{\cmd_5; \cmd_6} \To \configTwo{\Mem_3}{\cmd_5';\cmd_6} $:
		Command $ \cmd_2 $ must have the form of $ \cmd_7; \cmd_8 $ where $ \cmd_5 \eq \cmd_7, \cmd_6 \eq \cmd_8 $. We can infer from $ \cmd_5 \eq \cmd_7 $ and the evaluation assumption $  \configTwo{\Mem_1}{\cmd_5} \To \configTwo{\Mem_3}{\cmd_5'}  $ that $ \configTwo{\Mem_2}{\cmd_7} \To^* \configTwo{\Mem_5}{\cmd_7'} $ and $\Mem_5\eq \Mem_3$ and $\cmd_7'\eq \cmd_5'$. Thus, we can derive $ \configTwo{\Mem_2}{\cmd_7;\cmd_8} \To^* \configTwo{\Mem_5}{\cmd_7';\cmd_8} $.
		Therefore, we can choose $\cmd_4=\cmd_7';\cmd_8$ and $\Mem_4 = \Mem_5$.
		
		\item Case $ \configTwo{\Mem_1}{\assigneta{x}{e}} \To \configTwo{\switch(\Mem_1', x, \eta)}{\Skip}  $:
		From $ \cmd_1 \eq \cmd_2 $, we know $ \cmd_2  = (\assigneta{x}{e}) $, and $ \configTwo{\Mem_2}{\assigneta{x}{e}} \To \configTwo{\switch(\Mem_2', x, \eta)}{\Skip}  $. Trivially, $\Skip\eq \Skip$. Next, we show $\switch(\Mem_1', x, \eta)\eq \switch(\Mem_2', x, \eta)$.

From Lemma \ref{lem:expeq}, we know that $ \configTwo{\Mem_1}{e} \evalto v_1 $, $ \configTwo{\Mem_2}{e} \evalto v_2 $, then $ v_1 \eq v_2 $. We know $ \Mem_1' =  \Mem_1\Mupdate{x}{v_1} $ and $ \Mem_2' =  \Mem_2\Mupdate{x}{v_2} $. Thus, $ \Mem_1' \eq \Mem_2' $ given $ \Mem_1 \eq \Mem_2 $. Since there is no self-dependence, we have $\Mem_3(x)=\Mem_1'(x)\eq \Mem_2'(x)=\Mem_4(x)$. Next, we show for any $ \varx\not= x $, we have $ \Mem_3(\varx) \eq \Mem_4(\varx) $:
		\begin{itemize}
			\item Case $ x \not \in \free(\varx) $: by definition of $ \switch $, we have $ \Mem_3(\varx) = \Mem_1'(\varx)\eq \Mem_2'(\varx)=\Mem_4(\varx)$.
			\item Case $ x \in \free(\varx) $: by definition of $ \switch $, $ \varx $ is erased to $ 0 $ or $ [0] $ according to its type $ \labelof{\varx}{\Mem_1'} $ or $ \labelof{\varx}{\Mem_2'} $. 
			\begin{itemize}
				\item Case S-Asgn1, we have $\configTwo{\Mem_1}{e}\evalto n $, $\labelof{x}{\Mem_1}\LEQ L$ and $\Mem_1'(x)=n$. By Lemma~\ref{lem:expeq} and~\ref{lowexp}, $\configTwo{\Mem_2}{e}\evalto n $, $\labelof{x}{\Mem_2}\LEQ L$. So S-Asgn1 applies under $\Mem_2$ as well, and $\Mem_2'(x)=n ∧ \Mem_1'\eq \Mem_2'$. When $x'$ depends on no variable whose level is low under $\Mem_1$, we have $ \labelof{\varx}{\Mem_1'} = \labelof{\varx}{\Mem_2'} $ since all variables it depends on must be identical under $\Mem_1'$ and $\Mem_2'$. So $ \varx $ will be erased to the same value in this case. Otherwise, say $x'$ depends on $y$ such that $\labelof{y}{\Mem_1'}\not \LEQ L$. By Lemma~\ref{lowexp}, $y$ has a bracketed value under $\Mem_1'$. Since $\Mem_1'\eq \Mem_2'$, $y$ has a bracketed value under $\Mem_2'$ as well. By Lemma~\ref{lowexp} again, $\labelof{y}{\Mem_2'}\not \LEQ L$. Since $\G(y)\LEQ \G(x')$ when $x'$ depends on $y$, we have $\labelof{x'}{\Mem_1'}\not \LEQ L$ and $\labelof{x'}{\Mem_2'}\not \LEQ L$ by Lemma~\ref{lem:conserve}. Hence, $ \varx $ will be erased to [0] under $\Mem_1'$ and $\Mem_2'$. 
 
				\item Case S-Asgn2: we have $ \labelof{x}{\Mem_1} \not \LEQ L $. Since $\Mem_1\eq \Mem_2$, we have $\labelof{x}{\Mem_2} \not \LEQ L $ by Lemma~\ref{lem:expeq} and~\ref{lowexp}. So S-Asgn2 applies under $\Mem_2$ as well. Since there is no self dependence, we have $ \labelof{x}{\Mem_1'} = \labelof{x}{\Mem_1}\not \LEQ L$ and $ \labelof{x}{\Mem_2'} = \labelof{x}{\Mem_2}\not \LEQ L$. Since $x'$ depends on $x$, we must have $ \labelof{\varx}{\Mem_1'} \not \absleq L $ and $ \labelof{\varx}{\Mem_2'} \not \absleq L $. So by the definition of $ \switch $, $ \Mem_3(\varx) = \Mem_4(\varx) = [0] $.

				\item Case S-Asgn3: we have $ \configTwo{\Mem_1}{e}\evalto [n_1]$ for some $n_1$. By Lemma~\ref{lem:expeq}, $v_2 = [n_2]$ for some $n_2$. So S-Asgn3 applies under $\Mem_2$ as well. Moreover, given $ \G \proves e : \tau$, $\valueof{\tau}{\Mem_1}\not \LEQ L$ and $\valueof{\tau}{\Mem_2}\not \LEQ L$ by Lemma~\ref{lowexp}. We also know from typing rule T-Assgn that $\valid { \hypo (\before \eta) \sat \tau \join \pc \sqsubseteq \G(x)} $. Due to the correctness of predicate generation, $\tau \join \pc \LEQ \G(x)$. By Lemma~\ref{lem:conserve}, $\labelof{x}{\Mem_1}\not\LEQ L ∧ \labelof{x}{\Mem_2}\not\LEQ L$. Similar to case S-Asgn2, we know that $ \Mem_3(\varx) = \Mem_4(\varx) = [0] $ in this case.

			\end{itemize} 
		\end{itemize}
		
		\item Case $ \configTwo{\Mem_1}{\ifcmd{e}{\cmd_5}{\cmd_6}} \To\configTwo{\Mem_1}{\cmd_5} $ and $ \configTwo{\Mem_1}{\ifcmd{e}{\cmd_5}{\cmd_6}} \To\configTwo{\Mem_1}{\cmd_6} $:
		From $ \cmd_1 \eq \cmd_2 $, we know $ \cmd_2 $ must be $ \ifcmd{e}{\cmd_5}{\cmd_6} $. This rule is applied only when $e$'s value under $\Mem_1$ is not bracketed. By Lemma~\ref{lem:expeq}, $e$'s value is not bracketed under $\Mem_2$ and $e$ must evaluate to the same value under $\Mem_1$ and $\Mem_2$. Therefore, $ \cmd_1 $ and $ \cmd_2 $ must evaluate using the same rule. We can construct $c_4$ as the corresponding branch taken under $\Mem_1$.
		
		\item Case $ \configTwo{\Mem_1}{\ifcmd{e}{\cmd_5}{\cmd_6}} \To\configTwo{\Mem_1}{[\cmd_5]} $ and $ \configTwo{\Mem_1}{\ifcmd{e}{\cmd_5}{\cmd_6}} \To\configTwo{\Mem_1}{[\cmd_6]} $.
		From $\cmd_1\eq \cmd_2$, we know $ \cmd_2 $ must be $ \ifcmd{e}{\cmdt_5}{\cmdt_6}$. 
		This rule is applied when $ \configTwo{\Mem_1}{e} \evalto [n_1] $. By Lemma \ref{lem:expeq}, we know that $ \configTwo{\Mem_2}{e} \evalto [n_2] $ . We construct $\cmd_4$ as the branch taken under $\Mem_2$ in one step.
 According to the semantics, $\cmd_4$ evaluates to either $[\cmd_5]$ or $[\cmd_6]$. That is, both $ \cmd_3 $ and $\cmd_4$ evaluate to bracket commands. Hence, we have $ \cmd_3 \eq \cmd_4 $.
		
		
		\item Case $ \configTwo{\Mem_1}{\while{e}{\cmd_5}} \To\configTwo{\Mem_1}{\ifcmd{e}{(\while{e}{\cmd_5})}{\Skip}} $:
		From $ \cmd_1 \eq \cmd_2 $, we know $ \cmd_2 $ has the form of $ \while{e}{\cmd_5}$.
		Therefore, we construct $\cmd_4$ as \\ $\ifcmd{e}{(\while{e}{\cmd_5})}{\Skip}$. Hence, $ \Mem_2 \eq \Mem_4 $ and $ \cmdt_3 \eq \cmdt_4 $.
		
	\end{itemize}
\end{proof}

\begin{Theorem}[Soundness under Erasure Semantics] Any transforms program
that type checks satisfies noninterference:
\label{thm:erasuresoundness}
\begin{multline*}
	\forall \cmdt, \Mt_1, \Mt_2, \Mt_3, \Mt_4, \Gt, \ell, \actset'~. \\
	\Gt \proves \cmdt ∧
        \Mt_1 \lbleq{\Gt} \Mt_2 ∧ \\
	\configTwo{\Mt_1}{c} \ERTo{\actset'}^* \configTwo{\Mt_3}{\Skip} \AND
	\configTwo{\Mt_2}{c} \ERTo{\actset}'^* \configTwo{\Mt_4}{\Skip} \\
	\implies \Mt_1' \lbleq{\Gt} \Mt_2'
\end{multline*}
\end{Theorem}

\begin{proof}
By the construction of the extended language, for any particular $\ell$, the
relation $\Mt_1 \lbleq{\Gt} \Mt_2$ is the same as $\proves \Mt_1$, $\proves
\Mt_2$, $\Mt_1\eq \Mt_2$ in the extended language. We proceed by induction on
the number of steps in the execution under $\Mt_1$.

Case zero step: trivial since $c$ must be $\Skip$.

Case $N+1$ steps: consider the first step of the execution under $\Mt_1$. By
Lemma~\ref{lem:unwinding}, we know either $c$ diverges under $\Mt_2$ or we can
make multiple steps under $\Mt_2$ and $\Mt_3\eq \Mt_4$. By
Lemma~\ref{lem:preservation}, we also have $\proves \Mt_3$, $\proves \Mt_4$ and
the remaining programs $\cmdt_1$ and $\cmdt_2$ type-checks. Hence, result is
true by the induction hypothesis.
\end{proof}

\textbf{Proof of Theorem~\ref{thm:transsoundness}}
 \begin{multline*}
\forall \cmdt, \Mt_1, \Mt_2, \Mt_3, \Mt_4, \lbl, \actset, \actset', \Gt~. \\
         \configTwo{\actset}{\cmd}\tto \configTwo{\actset'}{\cmdt} ∧ \proves \Gt ∧
        \Gt \proves \cmdt~  \AND \Mt_1^{\actset} \lbleq{\Gt^{\actset}} \Mt_2^{\actset} \\
	\AND \configTwo{\Mt_1}{\cmdt} \To^* \configTwo{\Mt_3}{\Skip} \AND
 	\configTwo{\Mt_2}{\cmdt} \To^* \configTwo{\Mt_4}{\Skip} \\
	\implies \Mt_3^{\actset'} \lbleq{\Gt^{\actset'}} \Mt_4^{\actset'}
 \end{multline*}

\begin{proof}
Trivial by Theorem~\ref{thm:erasuresoundness} and
Lemma~\ref{lem:erasurecorrectness}, which states that the erasure semantics
agrees with the standard semantics on active copies in $\actset'$.
\end{proof}

\textbf{Proof of Theorem~\ref{thm:soundness}}
\begin{multline*}
	\forall \cmd, \cmdt, \Mem_1, \Mem_2, \Mem_1', \Mem_2', \lbl, \Gt, \actset, \actset'~. \\
	\configTwo{\actset}{\cmd}\tto \configTwo{\actset'}{\cmdt} ∧\proves \Gt ∧ \Gt \proves \cmdt ∧
        \Mem_1\lbleq{\Gt^{\actset}} \Mem_2 ∧ \\
	\configTwo{\Mem_1}{c} \To^* \configTwo{\Mem_1'}{\Skip} \AND
	\configTwo{\Mem_2}{c} \To^* \configTwo{\Mem_2'}{\Skip} \\
	\implies \Mem_1' \lbleq{\Gt^{\actset'}} \Mem_2'
\end{multline*}
\begin{proof}
Trivial by Theorem~\ref{thm:transsoundness} and Theorem~\ref{thm:correctness},
the correctness of program transformation. 
\end{proof}

\clearpage
\section{Enabling Flow-Sensitivity with Program Transformation}
\label{appendix:construct}

To facilitate the proof, we say a typing environment $\G'$ is an
\emph{extension} of $\G$, written $\G \ext \G'$, if $∀x\in
\Domain(\G).~\G(x)=\G'(x)$. Easy to check that this relation is an partial
order on environments (i.e., the relation satisfies reflexivity, antisymmetry
and transitivity). 



\begin{Lemma}
\label{lem:monotone}
\[\configTwo{\actset}{\bkcmd{\cmd}}\tto \configTwo{\actset'}{\cmdt} ⇒
\actset' ⊆ ({\actset}∪\FVars(\cmdt))\]
\end{Lemma}
\begin{proof}
By induction on the structure of $\cmd$.
\begin{itemize}

\item $\Skip$: 
$\actset'⊆\actset∪\FVars(\cmdt)$ since $\actset=\actset'$ in this case.

\item $\assign{x}{e}$:  we have
${\actset'}⊆{\actset}∪\FVars(\cmdt)$ since
$\actset'=\{x_i\}∪(\actset-\{\actset(x)\})$.

\item $\cmd_1;\cmd_2$: by the transformation rule we have
$\configTwo{\actset}{\bkcmd{\cmd_1}}\tto \configTwo{\actset_1}{\cmdt_1}$ and
$\configTwo{\actset_1}{\bkcmd{\cmd_2}}\tto \configTwo{\actset'}{\cmdt_2}$ for
some $\actset_1$. By the induction hypothesis, we have 
\[
 \actset_1 ⊆ ({\actset}∪\FVars(\cmdt_1)) ∧
 \actset' ⊆ ({\actset_1}∪\FVars(\cmdt_2))
\]
Hence, we have
\[\actset' ⊆ ({\actset_1}∪\FVars(\cmdt_1)∪\FVars(\cmdt_2)) ⊆
({\actset}∪\FVars(\cmdt_1)∪\FVars(\cmdt_2))\] Therefore, $\actset' ⊆
({\actset}∪\FVars(\cmdt_1;\cmdt_2))$.

\item $\ifcmd{e}{c_1}{c_2}$: by the transformation rule, we have
$\configTwo{\actset}{e}\!\tto\! \et$ and \\$\configTwo{\actset}{\bkcmd{\cmd_i}}\!\tto\!
\configTwo{\actset_i}{\cmdt_i}$, $i\in \{1,2\}$. By the induction hypothesis, we have
\[
 \actset_1 ⊆ ({\actset}∪\FVars(\cmdt_1)) ∧
 \actset_2 ⊆ ({\actset}∪\FVars(\cmdt_2))
\]

The transformed branch has the
form of $\cmdt_i;\actset_3:=\actset_i$ where
$\actset_3=\Phi(\actset_1,\actset_2)$. By the definition of $\Phi$,
$\actset_3⊆\actset_1∪\actset_2∪\FVars(\actset_3:=\actset_1;\actset_3:=\actset_2)⊆\actset∪\FVars(\cmdt)$. 

\item $\while{\expr}{\cmd}$: by the transformation rule, we have
\[
\inferrule
{ \configTwo{\actset}{\bkcmd{\cmd}} \tto \configTwo{\actset_1}{\cmdt_1} 
 	\quad \configTwo{\actset_1}{\bkcmd{\cmd}} \tto \configTwo{\actset_2}{\cmdt'} 
	\quad \configTwo{\actset_1}{e} \tto \et  }
{ \configTwo{\actset}{\bkcmd{\while{e}{\cmd}}} \tto \configTwo{\actset_1}{\phiassigneta{\actset_1}{\actset};\while{\et}{(\cmdt';\actset_1:=\actset_2)}} }
\]

By induction on $\configTwo{\actset}{\bkcmd{\cmd}} \tto
\configTwo{\actset_1}{\cmdt_1}$, we have $\actset_1⊆\actset∪\FVars(\cmdt_1)$.
For a variable $v$ in $\actset_1$ but not in $\actset$, we know that
$v\in \FVars(\actset_1:=\actset)$ since by definition, $\actset_1:=\actset$
assigns to any $v\in \actset_1 ∧ v\not\in \actset$.
Hence, $\actset_1⊆\actset∪\FVars(\cmdt)$. 
\end{itemize}
\end{proof}

The next lemma shows that the extension of a typing environment $\G$ subsumes
$\G$:

\begin{Lemma}
\label{lem:gextension}
\[\G\ext \G' ∧ \G \proves \cmd ⇒ \G' \proves \cmd \]
\end{Lemma}
\begin{proof}
We first show $\G\ext \G' ∧ \G \proves \expr:\tau ⇒ \G' \proves \expr:\tau$ by
induction on the structure of $\expr$. Then we can prove the lemma by induction
on the structure of $\cmd$.
\end{proof}

\textbf{Proof of Lemma~\ref{lem:extension}}
\begin{multline*}
\pc \fsproves \G \{\cmd\} \G' ∧ \configTwo{\actset}{\bkcmd{\cmd}}\tto
\configTwo{\actset'}{\cmdt}⇒ \\ 
∀v\in \Vars.~(\actset(v)=\actset'(v))⇒(\G(v)=\G'(v))
\end{multline*}
\begin{proof}
By induction on the structure of $\cmd$.

\begin{itemize}
\item $\Skip$: trivial since $\G=\G'$.

\item $\assign{x}{e}$: when $v=x$, $\actset'(v)=x_i\not=\actset(x)$ since $x_i$
is fresh. So the result is trivially true. For other variables, $\G(v)=\G'(v)$
by the HS typing rule~\ruleref{HS-Assign}.

\item $\cmd_1;\cmd_2$: by the transformation rule and HS typing rule, we have
$\configTwo{\actset}{\bkcmd{\cmd_1}}\tto \configTwo{\actset_1}{\cmdt_1}$ and
$\configTwo{\actset_1}{\bkcmd{\cmd_2}}\tto \configTwo{\actset'}{\cmdt_2}$ for
some $\actset_1$, as well as $\pc \fsproves \G \{\cmd_1\} \G_1$ and $\pc
\fsproves \G_1 \{\cmd_2\} \G'$  for some $\G_1$. By Lemma~\ref{lem:monotone},
$\actset'⊆\actset_1∪\FVars(\cmdt_2)$. So when $\actset(v)=\actset'(v)$, it must
be true that $\actset_1(v)=\actset'(v)$ since otherwise, $\actset'(v)$ must be
a fresh variable in $\cmdt_2$, and hence, cannot appear in $\actset$.
Therefore, we have $\actset(v)=\actset'(v)=\actset_1(v)$. By the induction
hypothesis, it must be true that $\G(v)=\G_1(v)=\G'(v)$.

\item $\ifcmd{e}{c_1}{c_2}$: by the HS typing rule, $\pc \fsproves \G \{\cmd_1\}
\G_1 ∧ \pc \fsproves \G \{\cmd_2\} \G_2 ∧ \G'=\G_1\join\G_2$. By the transformation
rules, $\configTwo{\actset}{\bkcmd{\cmd_i}}\tto \configTwo{\actset_i}{\cmdt_i},
i\in \{1,2\}$. 

By Lemma~\ref{lem:monotone}, $\actset_1⊆\actset∪\FVars(\cmdt_1)$. So when
$\actset(v)\not=\actset_1(v)$, $\actset_1(v)$ must be a fresh variable
generated in $\cmdt_1$, and hence, cannot be in $\actset_2$.  By the definition
of $\Phi$, $\actset_3(v)$ must be fresh as well. This contradicts the
assumption that $\actset(v)=\actset'(v)$. Hence, we have
$\actset(v)=\actset_1(v)$ (and similarly, $\actset(v)=\actset_2(v)$).
So $\G(v)=\G_i(v)$ by the induction hypothesis. Therefore,
$\G'(v)=\G_1(v)\join\G_2(v)=\G(v)$.

\item $\while{\expr}{\cmd}$: By rule~\ruleref{TRSF-While}, we have
$\configTwo{\actset}{\bkcmd{\cmd}}\tto \configTwo{\actset_1}{\cmdt_1}$,
where $\actset'$ is $\actset_1$ in this case. Hence, by the assumption, 
we have 
$\actset(v)=\actset_1(v)$.  By rule~\ruleref{HS-While}, there is a sequence of
environments $\G'_{i}, \G''_{i}$ such that $\pc\join \tau_i \proves \G'_i \{c\}
\G''_i$. By the induction hypothesis, $\G''_i(v)=\G'_i(v)$. Since
$\G'_0=\G$ and $\G'_{i+1}=\G\join \G''_i$ in rule~\ruleref{HS-While}, we can
further infer that $\G_{i+1}'(v)=\G_i''(v)$.  Hence, we have
$\G'(v)=\G_n(v)=\G_0(v)=\G(v)$.

\end{itemize}

\end{proof}

\begin{Lemma}
\label{lem:fsexpr}
\begin{multline*}
\G \fsproves \expr:\tau ∧ \configTwo{\actset}{\expr}\tto \et ∧ \G = \Gt^{\actset} ⇒ \\ 
\Gt \proves \et:\tau
\end{multline*}
\end{Lemma}
\begin{proof}
By rule induction on the transformation.
\end{proof}

\textbf{Proof of Theorem~\ref{thm:constructcorrect}} We prove a lightly
stronger version of Theorem~\ref{thm:constructcorrect}:
\begin{multline*}
∀\cmd, \cmdt,\pc, \actset, \actset', \G, \G', \Gt.~\\
(\pc, \G, \actset) \{\bkcmd{\cmd}\tto \cmdt\} (\G', \actset') \produce \Gt \\
⇒ \Domain(\Gt)⊆\actset∪\FVars(\cmdt) ∧ \G = \Gt^{\actset} ∧ \G' = \Gt^{\actset'} ∧ \Gt, \pc \proves \cmdt 
\end{multline*}
\begin{proof}
By induction on the structure of $\cmd$.

\begin{itemize}
\item $\Skip$: $\Gt, \pc \proves \cmdt$ is trivial since $\Skip$ can be
type-checked with any $\pc,\Gt$. Other conditions are trivial.

\item $\assign{x}{\expr}$: by the construction rule, we have $\G'(x)=\tau$ for some $\tau$,
$\cmdt$ has the form of $x_i:=\et$, $\actset'(x)=x_i$ and
$\Gt=\G_{\actset}∪\{x_i\mapsto \tau\}$.

\begin{itemize}
\item $\G = \Gt^{\actset}, \G' = \Gt^{\actset'}$:
Since $x_i$ is fresh, $\G = \Gt^{\actset}$. Moreover,
$\Gt^{\actset'}=\G\{x\mapsto \tau\}=\G'$.

\item $\Gt, \pc \proves \cmdt$: let $\G \fsproves e:\tau_0$. By the HS typing rule,
we have $\tau = \tau_0\join \pc$. Moreover, we have $\Gt \proves \et:\tau$ due
to Lemma~\ref{lem:fsexpr} and the fact $\G= \Gt^{\actset}$. By the construction,
$\Gt(x_i)=\tau$. Hence, $\Gt,\pc\proves x_i := \et$.

\item $\Domain(\Gt)⊆\actset∪\FVars(\cmdt)$: By the construction,
$\Domain(\Gt)=\actset∪\{x_i\}$.
\end{itemize}

\item $\cmd_1;\cmd_2$: from the construction rule, we have $(\pc, \G, \actset)
\{ \bkcmd{c_1}\tto \cmdt_1 \} (\G'',\actset'') \produce \Gt_1$ and\\ 
$(\pc, \G'', \actset'') \{ \bkcmd{c_2} \tto \cmdt_2 \}
(\G', \actset') \produce \Gt_2 $ for some $\G''$,$\actset''$, $\Gt_1$ and $\Gt_2$. 
By the induction hypothesis, we have 
\[\G= \Gt_1^{\actset} ∧ \G''= \Gt_1^{\actset''} ∧ \Domain(\Gt_1)⊆\actset∪\FVars(\cmdt_1)\]
\[\G''= \Gt_2^{\actset''} ∧ \G'= \Gt_2^{\actset'} ∧ \Domain(\Gt_2)⊆\actset''∪\FVars(\cmdt_2)\]
\[\Gt_1,\pc
\proves \cmdt_1 ∧ \Gt_2,\pc \proves \cmdt_2\]

\begin{itemize}
\item $\G=\Gt^{\actset},\G'=\Gt^{\actset'}$: By the construction, we have
$\Gt_1\ext \Gt$. Next, we check $\Gt_2\ext \Gt$.

Due to the results above on $\Domain(\Gt_1)$ and $\Domain(\Gt_2)$, we know that
if $\vt \in \Domain(\Gt_1)∩ \Domain(\Gt_2)$, then $\vt\in \actset''$ since
$\FVars(\cmdt_2)$ contain fresh variables generated in $\cmdt_2$ (hence, not in $\actset$ and $\cmdt_1$).  Therefore,
$\Gt_1(\vt)=\Gt_2(\vt)$ because $\Gt_1^{\actset''}=\G''=\Gt_2^{\actset''}$ by the induction hypothesis.
Therefore, $\Gt_2\ext \Gt$.

Moreover, by the definition, $\G=\Gt_1^{\actset}$ is equivalent to
$\G_{\actset}\ext \Gt_1$. So we have $\G_{\actset}\ext \Gt$, and hence,
$\G=\Gt^{\actset}$.  Similarly, we can prove that $\G'=\Gt^{\actset'}$.

\item $\Gt, \pc \proves \cmdt$: we know that $\Gt,\pc \proves \cmdt_1 ∧ \Gt,\pc
\proves \cmdt_2$ by Lemma~\ref{lem:gextension}. So $\Gt,\pc \proves
\cmdt_1;\cmdt_2$. 

\item $\Domain(\Gt)⊆\actset∪\FVars(\cmdt)$: by the construction,
$\Domain(\Gt)=\Domain(\Gt_1)∪\Domain(\Gt_2)$. The result is true since
$\actset''⊆\actset∪\FVars(\cmdt_1)$ by Lemma~\ref{lem:consistency} and
Lemma~\ref{lem:monotone}.

\end{itemize}

\item $\ifcmd{\expr}{c_1}{c_2}$: from the construction rule, we have $(\pc\join
\tau, \G, \actset) \{ \bkcmd{c_i}\tto \cmdt_i \} (\G_i, \actset_i) \produce \Gt_i$  for $i\in
\{1,2\}$, and $\Phi(\actset_1,\actset_2)\tto \actset_3$, where $Γ \fsproves e:\tau$.  By the induction
hypothesis, we have 
\[\G = \Gt_i^{\actset} ∧ \G_i = \Gt_i^{\actset_i} ∧ \Gt_i,\pc\join \tau \proves
\cmdt_i ∧ \Domain(\Gt_i)⊆\actset∪\FVars(\cmdt_i)\]

\begin{itemize}
\item $\Gt_1\ext \Gt,\Gt_2\ext \Gt$: By the construction, we have $\Gt_1\ext
\Gt$.  Next, we check $\Gt_2\ext \Gt$.

Since $\Domain(\Gt_i)⊆\actset∪\FVars(\cmdt_i)$, if $\vt\in
\Domain(\Gt_1)∩\Domain(\Gt_2)$, then it must be true that $\vt\in \actset$.
Since we know that $\G=\Gt_1^{\actset}$ and $\G=\Gt_2^{\actset}$ from the
induction hypothesis, $\Gt(\vt)=\Gt_1(\vt)=\Gt_2(\vt)$ for such variables.
Hence, we have $\Gt_2\ext \Gt$. 

\item $\G=\Gt^{\actset},\G'=\Gt^{\actset_3}$:
We have $\G=\Gt^{\actset}$ since $\G=\Gt_1^{\actset}$ and $\Gt_1\ext \Gt$. To
check $\G'=\Gt^{\actset_3}$, consider $\vt\in \actset_3∩(\Domain(\Gt_1) ∪
\Domain(\Gt_2))$. We use $v$ to denote $\proj{\vt}$.

\begin{itemize}
\item When $\vt \in \Domain(\Gt_1)$: we have $\vt\in \actset$ or $\vt\in
\FVars(\cmdt_1)$ since $\Domain(\Gt_1)⊆\actset∪\FVars(\cmdt_1)$. 

In the latter case, we have $\actset_1(v)\not=\actset_2(v)$ since
$\actset_2⊆\actset∪\FVars(\cmdt_2)$ by Lemma~\ref{lem:consistency} and
Lemma~\ref{lem:monotone}. So $\actset_3(v)$ must be a
fresh variable by the definition of $\Phi$, and hence $\vt\not\in \actset_3$.
Contradiction. 

So $\vt\in \actset$. By the assumption that $\vt \in \actset_3$ and the
definition of the $\Phi$ function, we have $\vt\in \actset_1$ and $\vt \in
\actset_2$. That is,
$\actset(v)=\actset_1(v)=\actset_2(v)$.  By Lemma~\ref{lem:extension}, we have
$\G_1(v)=\G(v)=\G_2(v)$. Hence, $\G'(v)=\G_1(v)\join \G_2(v)=\G(v)=\Gt(\vt)$,
where the last equation is true since $\vt\in \actset ∧ \Gt^{\actset}=\G$.

\item When $\vt \in \Domain(\Gt_2)$: same argument as the case above.
\end{itemize}

Therefore, for any $\vt\in \actset_3∩(\Domain(\Gt_1) ∪ \Domain(\Gt_2))$,
$\Gt(\vt)=\G'(v)$. For other variables, $\Gt(\vt)=\G'(v)$ is
trivial from the construction. So $\G'=\Gt^{\actset_3}$.

\item $\Gt,\pc \proves \cmdt$:
We know that $\configTwo{\actset}{\expr}\tto {\et}$ in the transformation
assumption. Given $\Gt^{\actset}=\G$, we have $\Gt\proves \et:\tau$ by
Lemma~\ref{lem:fsexpr}. Moreover, $\Gt,\pc\join \tau \proves \cmdt_i$ since
$\Gt_1\ext \Gt$, $\Gt_2\ext \Gt$. Furthermore, both
$\assign{\actset_3}{\actset_1}$ and $\assign{\actset_3}{\actset_2}$ type-checks
by our construction since: 1) by the defintion, the extra assignments only
assign to fresh variables in $\actset_3$ (i.e., $\vt\not\in
\actset_1∩\actset_2$); 2) for those variables, we have
$\Gt(\vt)=\G'(v)$ by the construction; 3) the HS system ensures that
$\G_i(v)\LEQ \G'(v)$ for $i\in\{1,2\}$ and $\pc\join\tau \LEQ
\G_i(v)$ for $i=1$ or $2$ ($v$ must be assigned to under some branch since
$\vt\not\in \actset_1∩\actset_2$). Therefore, we have 
$\pc\join\tau\LEQ \G'(v)=\Gt(\vt)$, and $\Gt(\actset_i(v))=\G_i(v) \LEQ \G'(v)= \Gt(\vt)$.

\item $\Domain(\Gt)⊆\actset∪\FVars(\cmdt)$: by the construction,
$\Domain(\Gt)=\Domain(\Gt_1)∪\Domain(\Gt_2)∪\actset_3$. By
Lemma~\ref{lem:monotone}, $\actset_3⊆\actset∪\FVars(\cmdt)$. Result is true
since we also have $\Domain(\Gt_i)⊆\actset∪\FVars(\cmdt_i)$.
\end{itemize}

\item $\while{e}{c}$: by the construction rule,  we have
\[\G'\fsproves \expr:\tau, (\tau \join \pc, \G', \actset_1) \{ \bkcmd{\cmd} \tto \cmdt'
\} (\G',\actset_2) \produce \Gt_0\]
By the induction hypothesis, we have 
\[\G' = \Gt_0^{\actset_1} ∧ \G' = \Gt_0^{\actset_2} ∧ \Gt_0, \pc\join \tau
\proves \cmdt' ∧ \Domain(\Gt_0)⊆\actset_1∪\FVars(\cmdt')\]

\begin{itemize}

\item $\G=\Gt^{\actset}$, $\G'=\Gt^{\actset_1}$:
We have $\Gt_0\ext \Gt$ by the construction. Given $\G'=\Gt_0^{\actset_1}$, we know
that $\G'=\Gt^{\actset_1}$. Next, we show that $\G=\Gt^{\actset}$. 

\begin{itemize}
\item When $\vt\in \actset∩\Domain(\Gt_0)$ (we use $v$ to denote $\proj{\vt}$).
Since $\Domain(\Gt_0)⊆\actset_1∪\FVars(\cmdt')$ from the induction hypothesis,
we have $\vt\in \actset_1$ or $\vt$ is fresh in $\cmdt'$. But the latter must
be false since $\vt\in \actset$. So $\vt \in \actset_1$.

From the construction rule assumption, we have $\pc\fsproves
\G\{\while{e}{c}\}\G'$. By Lemma~\ref{lem:consistency}, we have
$\configTwo{\actset}{\bkcmd{\while{e}{c}}}\tto \configTwo{\actset_1}{\cmdt}$.
In addition, we just showed that $\vt\in \actset$ and $\vt\in \actset_1$.
Hence, by Lemma~\ref{lem:extension}, $\G(v)=\G'(v)$.

Because $\vt\in \actset_1$ and $\vt\in \actset$, $\actset(v)=\actset_1(v)$.
Therefore, $\Gt^{\actset}(v)=\Gt(\actset(v))=\Gt(\actset_1(v))=\G'(v)
\text{(since $\G'=\Gt^{\actset_1}$)}=\G(v)$.  

\item When $\vt\in \actset-\Domain(\Gt_0)$, the result is trivial by
the construction. 
\end{itemize}

\item $\Gt,\pc \proves \cmdt$: we have showed that $\Gt^{\actset_1}=\G'$.  In the transformation
assumption, we have $\configTwo{\actset_1}{\expr}\tto \et$. So by
Lemma~\ref{lem:fsexpr}, $\Gt\proves \et:\tau$. Moreover, we have $\Gt, \pc\join
\tau\proves \cmdt'$ by Lemma~\ref{lem:gextension} and the induction hypothesis
$\Gt_0, \pc\join \tau \proves \cmdt'$. Next, we show that the extra assignments
(i.e., $\actset_1:=\actset$ and $\actset_1:=\actset_2$) type-check.

Since $\G' = \Gt_0^{\actset_1} ∧ \G' = \Gt_0^{\actset_2}$ by the induction
hypothesis, we have $\Gt^{\actset_1}=\Gt^{\actset_2}$. Moreover, by the same
argument as in the ``if'' case, the LHS of $\assign{\actset_1}{\actset_2}$ must
have a level that is higher than $\pc\join \tau$ in the HS system.  So
$\Gt,\pc\join \tau \proves \assign{\actset_1}{\actset_2}$ . Moreover, we know
that $∀v\in \Vars.~\G(v)\LEQ \G'(v)$ in the HS system. Since $\Gt^{\actset}=\G$
and $\Gt^{\actset_1}=\G'$, we have $\Gt,\pc \proves \assign{\actset_1}{\actset}$. 

\item $\Domain(\Gt)⊆\actset∪\FVars(\cmdt)$: by the construction,
$\Domain(\Gt)=\Domain(\Gt_0)∪\actset$. From the induction hypothesis, we have
$\Domain(\Gt_0)⊆\actset_1∪\FVars(\cmdt')$. Hence,
$\Domain(\Gt_0)⊆\actset_1∪\FVars(\cmdt)$.
We know from Lemmas~\ref{lem:consistency} and~\ref{lem:monotone} that
$\actset_1⊆\actset∪\FVars(\cmdt)$. Hence,
$\Domain(\Gt)⊆\actset∪\FVars(\cmdt)$.
\end{itemize}

\end{itemize}
\end{proof}

\fi
\end{document}